\documentclass{jfm}

\usepackage[latin1]{inputenc}
\usepackage[english]{babel}
\usepackage{float}
\usepackage{textcomp}
\usepackage{amsmath}
\usepackage{graphicx}
\usepackage{setspace}
\usepackage{esint}
\usepackage{natbib}

\usepackage{psfrag}

\usepackage[T1]{fontenc}               
\usepackage[section]{placeins}         

\usepackage{fancyhdr}
\usepackage{mathrsfs}
\usepackage{amsfonts}
\usepackage{lscape}
\usepackage{longtable}

\usepackage{array}
\newcolumntype{L}[1]{>{\raggedright\let\newline\\\arraybackslash\hspace{0pt}}m{#1}}
\newcolumntype{C}[1]{>{\centering\let\newline\\\arraybackslash\hspace{0pt}}m{#1}}
\newcolumntype{R}[1]{>{\raggedleft\let\newline\\\arraybackslash\hspace{0pt}}m{#1}}

\newcommand{\lyxdot}{.}

\ifCUPmtlplainloaded \else
  \checkfont{eurm10}
  \iffontfound
    \IfFileExists{upmath.sty}
      {\typeout{^^JFound AMS Euler Roman fonts on the system,
                   using the 'upmath' package.^^J}%
       \usepackage{upmath}}
      {\typeout{^^JFound AMS Euler Roman fonts on the system, but you
                   dont seem to have the}%
       \typeout{'upmath' package installed. JFM.cls can take advantage
                 of these fonts,^^Jif you use 'upmath' package.^^J}%
      }
  \else
  \fi
\fi

\ifCUPmtlplainloaded \else
  \checkfont{msam10}
  \iffontfound
    \IfFileExists{amssymb.sty}
      {\typeout{^^JFound AMS Symbol fonts on the system, using the
                'amssymb' package.^^J}%
       \usepackage{amssymb}%

      }{}
  \fi
\fi

\ifCUPmtlplainloaded \else
  \IfFileExists{amsbsy.sty}
    {\typeout{^^JFound the 'amsbsy' package on the system, using it.^^J}%
     \usepackage{amsbsy}}
    {}
\fi

\newsavebox{\astrutbox}
\sbox{\astrutbox}{\rule[-5pt]{0pt}{20pt}}

\title[Local properties of highly nonlinear unsteady waves]{On the local properties of highly nonlinear unsteady gravity water
waves. Part 1. Slowdown, kinematics and energetics}

\author[X. Barthelemy, M.L. Banner, W.L. Peirson, F. Dias, M. Allis]{X. Barthelemy$^1,^2$ \thanks{Email address for correspondence: x.barthelemy@unsw.edu.au},%
 M.L. Banner$^1$,  W.L. Peirson$^2$,\break
   F. Dias$^3$ and M. Allis$^2$ }

\affiliation{$^1$School of Mathematics and Statistics, UNSW Australia, Sydney NSW 2052, Australia\\[\affilskip]
$^2$Water Research Laboratory, School of Civil and Environmental Engineering, UNSW Australia, Sydney NSW 2052, Australia\\
$^3$UCD School of Mathematical Sciences, University College Dublin, Belfield, Dublin 4, Ireland}

\pubyear{2010}
\volume{650}
\pagerange{119--126}
\date{?; revised ?; accepted ?. - To be entered by editorial office}

\begin{document}
\maketitle 
\begin{abstract}
The kinematic properties of unsteady highly non-linear 3D wave groups have been investigated using a numerical wave tank. Although carrier
wave speeds based on zero-crossing analysis remain within $\pm 7\%$ of linear theory predictions, crests and troughs locally undertake
a systematic cyclical leaning from forward to backward as the crests/troughs transition through their maximum amplitude. Consequently, both crests
and troughs slow down by approximately $15\%$ of the linear velocity, in sharp contrast to the predictions of finite amplitude Stokes steady
wavetrain theory. Velocity profiles under the crest maximum have been investigated and surface values in excess of 1.8 times the equivalent Stokes velocity
can be observed. Equipartitioning between depth-integrated kinetic and potential energy holds globally on the scale of the wave group.
However, equipartitioning does not occur at crests and troughs (even for low amplitude Stokes waves), where the local ratio of potential
to total energy varies systemically as a function of wave steepness about a mean value of 0.67. 
\end{abstract}

\section{Introduction}

Wind-generated ocean waves at the air-sea interface propagate predominantly in group or packet structures. Group structure behaviour is not limited
to ocean waves, but occurs naturally in many systems. These unsteady wave groups can exhibit a complex nonlinear life cycle, especially
in focal zones where there is rapid wave energy concentration. Their structure and the associated carrier wave propagation properties of
dispersion, directionality and nonlinearity, and their interplay, represents a knowledge gap beyond present analytical treatment. Understanding
natural ocean wave propagation is not only relevant academically but is potentially important when studying ocean-atmosphere exchanges,
as well as the design of structures such as open ocean platforms. Crest speeds are a key consideration in the assessment of wave impact loading on structures, where
flow-induced forces are proportional to the square of the water velocity. In other applications, measurements of initial whitecap speeds have
been proposed as a method for inferring ocean wavelengths (\citet{Phillips1985}) assuming that the Stokes dispersion relationship is applicable. Any
systematic crest speed slowdown just prior to breaking onset will significantly alias wavelength determinations from the quadratic
relationship between Stokes wavelength and speed.

Stokes classical deep water wave theory ((\citet{Stokes1847}) was developed for a steady, uniform train of two-dimensional (2D) non-linear,
deep-water waves of small-to-intermediate mean steepness $ak\left(=2\pi\frac{a}{\lambda}\right)$, where $a$ is the wave amplitude and $\lambda$ is the wavelength. The wave speed c increases with $ak$ as found to 5th order in wave
steepness by \citet{Fenton1985}:

\begin{equation}
c=c_{0}\left(1+\frac{1}{2}\left(ak\right)^{2}+\frac{1}{8}\left(ak\right)^{4}\mbox{+higher order terms in \ensuremath{\left(ak\right)^{6+}}}\right)\label{eq:Stokes}
\end{equation}

where $c_{0}$ is the wave speed of linear (infinitesimally steep)
waves.

Increasing the steepness to the Stokes limit ($ak\sim0.42$) in eq
\eqref{eq:Stokes} gives the limiting speed of maximally-steep steady
waves of $1.1c_{0}$. Increased wave steepness has long been associated
with higher wave speeds. Historically, Stokes theory has been the
default theoretical approach for describing ocean waves (\citet{Kinsman1965}
 and \citet{Wiegel1964}). Contemporary examples are \citet{Beya2012}
who found close agreement between monochromatic wave near-surface
velocities and predictions based on 5th order Stokes theory. The observational
study of \citet{Grue2003} using PIV to measure velocities under extreme
and breaking crests showed that inviscid predictions are in good agreement
with measured particle velocities. \citet{Johannessen2010} showed
that at the crest, near-surface velocities agree
with second-order Stokes theory, which is capable of describing the
free surface kinematics very accurately provided that the local underlying
regime of free waves can be identified. A principal
constraint of the Stokes wavetrain framework is the imposed uniform
spatial and temporal periodicity, which provides the basis for the
extensive analytic treatment in the literature of this intrinsically
nonlinear free surface problem.

However, ocean waves propagate naturally in unsteady groups that evolve
dynamically. Relaxing the periodicity constraints allows additional
degrees of freedom to develop as shown in a suite of laboratory studies.
\citet{Melville1983} found a wave speed variation of between -17\%
and +32\% around the linear phase velocity. \citet{Shemer2013} demonstrated
that for a wave group with a broad spectral bandwidth, the crest propagation
velocity of the dominant waves differs significantly from both their
phase and group velocities. Deep-water breaking wave studies (e.g.
\citet{Rapp1990}, \citet{Stansell2002}, \citet{Jessup2005}, \citet{Melville2002}, \citet{MELVILLE2002a}) all found a significant (O(20\%)) breaking
crest speed slowdown relative to the expected linear phase velocity
(Eq. \ref{eq:Stokes}). For focusing wave packets, \citet{Johannessen2001}
and \citet{Johannessen2003} reported a slowdown in the expected crest
velocity at focus of around 10\%.

Understanding wave crest slowdown behaviour is central to both refining
present knowledge on water-wave propagation and dynamics, and effective
implementation of Phillips' spectral framework for breaking waves
(\citet{Phillips1985}, \citet{Gemmrich2008}, \citet{Kleiss2010a}).
The groupiness of natural ocean waves gives rise to a suite of differences
from Stokes waves predictions. \citet{Baldock1996} report an ensemble
of directional wave modes of different frequency components focusing
in space and time to produce an unsteady wave-group in the laboratory.
Surface elevations and subsurface particle kinematics were compared
with linear wave theory and the \citet{Longuet-Higgins1960a} second-order
solution of the wave-wave interactions. Their study shows that nonlinear
wave-wave interactions produce a highly nonlinear wave-group with
crests higher and troughs shallower than expected from numerical simulations
by \citet{Longuet-Higgins1987} and experiments by \citet{Miller1991}.
\citet{Sutherland1995} previously obtained very similar characterisations
although they also concluded that the wave kinematics were Stokes-like.
\citet{Song2002}, \citet{Banner2007} and \cite{Viotti2014} report complementary numerical
and experimental studies that demonstrate that intra-group and inter-wave
energy transfers cannot be neglected when characterising wave
group kinematics and dynamics.

In this contribution we show that Stokes characterisations have significant
limitations when applied to fully non-linear unsteady wave groups.
Specifically, important properties derived from \citet{Stokes1847}
theory and subsequent refinements (e.g. \citet{Fenton1985}) change
fundamentally when stationarity is relaxed. Under wave crest maxima,
the velocity profiles, energetics and energy partitioning all differ
systematically from those of a comparably steep Stokes wave. The unsteady
leaning of waves (\citet{Tayfun1986}) is shown to be a critical aspect,
determining the actual wave crest speed. Our preliminary numerical studies
of chirped wave packets have already been verified against observations
of wave groups in the laboratory and in the field (\citet{Banner2014}). \citet{Fedele2014} provided a theoretical explanation and further validation of the crest slowdown in terms of geometric phases.

\section{Numerical simulation}

For modulated waves, the nonlinear Schrödinger (NLS) equation and
its higher-order extensions (\citet{Zakharov1968}) have been widely
applied. NLS models have been used extensively to study the evolution
of quasi-monochromatic waves and instabilities due to resonant interactions
(e.g. the Benjamin\textendash Feir instability (\citet{Benjamin1967}).
Extensions of the NLS equation include the Zakharov equation which
has been applied to describe the long-time evolution of the spectrum
of weakly nonlinear, dispersive waves.

There has been a growing interest in the development of three-dimensional
models which inherently incorporate non-linearity and associated dispersion
effects. The broad-bandedness in both frequency and direction of real
sea states poses significant challenges in numerical simulation. \citet{Bateman2001}
demonstrated the importance of directionality and the consequent benefits
of efficient wave modelling during comparison of numerical simulations
with the laboratory observations of \citet{Johannessen2001}. High-order
spectral expansion approaches using efficient FFT solvers for application
to 3D waves have been developed (e.g. \citet{Ducrozet2011}, \cite{Dias2015}). Another
option is to solve the full Navier-Stokes equations (\citet{Park2003}),
but viscous flow solvers tend to be too dissipative and computationally
time-consuming.

Numerical models of potential 3D wave propagation can be divided into
three main categories: (a) boundary element integral methods (BEM):
e.g. \citet{Baker1982}, \citet{Bateman2001}, \citet{Clamond2001},
\citet{Grilli2001}, \citet{Fochesato2007}, \citet{Guyenne2006}, \citet{XUE2001}, \citet{Hou2002}, \citet{Fructus2005}; (b) finite
element method (FEM) e.g. \citet{Ma2001}; (c) spectral methods:
e.g. \citet{Dommermuth1987}, \citet{West1987}, \citet{Craig1993},
\citet{Nicholls1998}, \citet{Bateman2001}.

Spectral methods based on perturbation expansions are known to be
very efficient. These methods reduce the water wave problem from one
posed inside the entire fluid domain to one posed on the boundary
alone, thus reducing the dimension of the formulation. This reduction
can be accomplished by using integral equations over the boundary
of the domain (so-called boundary integral methods) or by introducing
boundary quantities which can be expanded as Taylor series for reference
domain geometries (\citet{Xu2009}). Both approaches
have been summarised recently by \citet{Ma2010}. BEM techniques are
efficient for representing wave propagation and overturning until
the wave surface reconnects (\citet{Grilli1996}).

The present study used a boundary element numerical wave tank code
called WSIM, which is a 3D extension of the 2D code developed by \citet{Grilli1989}
to solve the single-phase wave motions of a perfect fluid. It has
been applied extensively to the solution of finite amplitude wave
propagation and wave breaking problems (see chapter 3 of \citet{Ma2010}).
The ideal fluid assumption makes WSIM unable to simulate breaking
impact subsequent to surface reconnection. However, its potential
theory formulation enables it to simulate wave propagation in a CPU-efficient
way, without the diffusion issues of viscous numerical codes. The
simulation of wave generation and development of the onset of breaking
events can be carried out with great precision, as demonstrated by
\citet{Fochesato2006}, \citet{Fochesato2007}.

WSIM has been validated extensively and shows excellent energy conservation
(\citet{Grilli1989}, \citet{Grilli1990}, \citet{Grilli1994}, \citet{Grilli1996},
\citet{Grilli1997}, \citet{Grilli2001},
\citet{Fochesato2004}, \citet{Fochesato2006},
\citet{Fochesato2007}). Its kinematical accuracy has been demonstrated
by comparison with the analytical solutions for infinitesimal sine
waves found in \citet{Phillips1977}.

\subsection{Governing equations}

\begin{figure}
\psfrag{X}[][]  { $\mathbf{x}$}
\psfrag{Y}[][]  { $\mathbf{y}$}
\psfrag{Z}[][]  { $\mathbf{z}$}
\psfrag{xt}[][]  { $\mathbf{x(t)}$}
\psfrag{n}[][]  { $\vec{\mathbf{n}}$}
\psfrag{s}[][] { $\vec{\mathbf{s}}$}
\psfrag{m}[][] { $\vec{\mathbf{m}}$}
\psfrag{Gfs}[][] { $\mathbf{\Gamma_{FS}}$}
\psfrag{O}[][] { $\mathbf{O}$}
\psfrag{Beach}[][] { $\mathbf{Absorbing Piston}$}
\psfrag{Gamma}[][] { $\mathbf{\Gamma}$}
\psfrag{Paddle}[][] { $\mathbf{Paddle}$}
\centering
\includegraphics[height=6cm]{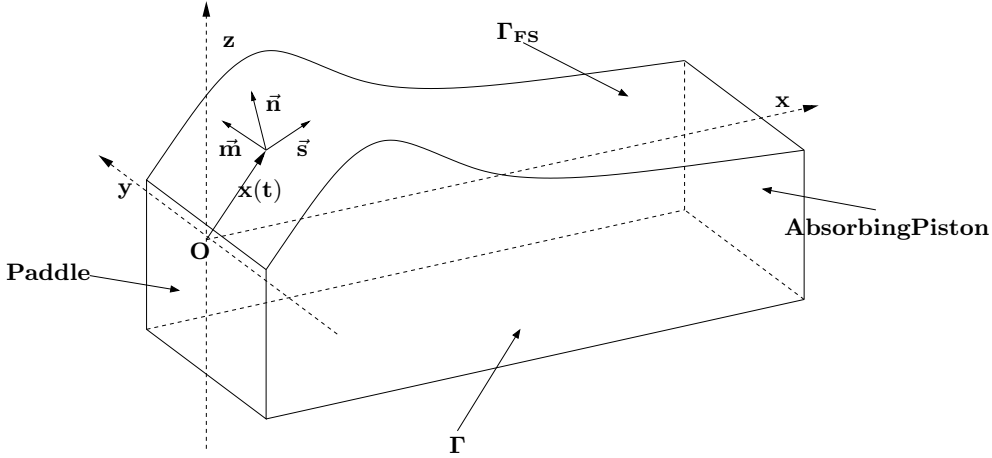}
\caption{Simulation domain}
\label{fig:Simulation-domain}
\end{figure}

WSIM uses a mixed Eulerian-Lagrangian time-updating scheme for irrotational
motion described by the velocity potential $\phi\left(\mathbf{x},t\right)$,
in a Cartesian coordinate system $\mathbf{x}=\left(x,y,z\right)$
with constant pressure at the open water surface. $z$ is the vertical
upward direction and $z=0$ the still water surface (Figure \ref{fig:Simulation-domain}).

Fluid velocity is defined as $\mathbf{u}=\mathbf{\nabla\phi}=\left(u,v,w\right)$.
The continuity equation leads to the Laplace equation for the potential
within the fluid domain $\Omega\left(t\right)$ (\ref{eq:laplace}).
The symbols $\Gamma$ and $\Gamma_{FS}$ are used to denote the entire
domain boundary and the free surface respectively.

\begin{eqnarray}
\nabla^{2}\phi & = & 0\textrm{, on }\Omega\label{eq:laplace}\\
\frac{D\mathbf{r}}{Dt} & = & \mathbf{u}=\mathbf{\nabla\phi}\textrm{, on }\Gamma_{FS}\label{eq:kinematic}\\
\frac{D\phi}{Dt} & = & -gz+\frac{1}{2}\nabla\phi\nabla\phi-\frac{P_{0}}{\rho}\textrm{, on }\Gamma_{FS}\label{eq:dynamic}\\
\partial_{n}\phi & = & 0\textrm{, on }\Gamma\backslash\Gamma_{FS}\label{eq:boundary}
\end{eqnarray}

where $\mathbf{r}$ is the position vector of a fluid particle on
the free surface, $g$ the gravitational acceleration, $P_{0}$ the
atmospheric pressure, $\rho$ the fluid density and $\frac{D}{Dt}\left(=\frac{\partial}{\partial t}+\nabla\phi\cdot\nabla\right)$ 
 the Lagrangian (or material) time derivative.

The free surface (domain $\Gamma_{FS}$) is described by fully-nonlinear
kinematic (eq \ref{eq:kinematic}) and dynamic (\ref{eq:dynamic})
equations. Recent developments have implemented a 3D snake wave paddle
at one vertical face of the domain which is described subsequently.
The far face of the numerical wave tank from the paddle has an absorbing
beach which damps any incident wave energy as described in \citet{Grilli1997}.
All remaining faces of the domain have a zero-flux boundary condition
($\Gamma\backslash\Gamma_{FS}$) (eq \ref{eq:boundary}).

\subsection{Discretisation}

\label{discretisation}

Boundary geometry and field variables are represented by 16-node quadrilateral elements, providing bi-cubic local interpolation of the solution between
the nodes. High-order tangential derivatives required for the time updating are calculated in a local curvilinear coordinate system,
using 25-node fourth-order quadrilateral elements to retain third order accuracy in the spatial derivative. This results in a high-order
integration scheme over the discretisation elements. The solver uses a fast PTMA scheme introduced by \citet{Fochesato2006}.

Once preliminary testing of WSIM was complete, two spatial resolutions were used during the present simulations to quantify sensitivity to
the discretisation. These were 8 points per wavelength in (so-called) medium resolution and 16 points per wavelength in high resolution
in the propogation ($x$) direction. The number of points in the transverse $y$ and depth $z$ directions was adjusted to keep element aspect ratios
close to 1.

Even if WSIM internally works with a different reference system, the results in this paper have been chosen to be expressed in deep water
units. Model time scales are non-dimensionalised by the base frequency $\omega_{p}$ of the generating paddle signal, and its corresponding length scale
is the deep water wavelength $\lambda_p$. These choices impose
a gravitational acceleration of $g_{DW}=2\pi$ and a reference linear velocity $C_{DW}=\frac{g_{DW}}{\omega_{DW}}=1$.

Non-dimensional lengths can be computed using $L_{DW}=\frac{L}{\lambda_p}$ and the time related quantities using $T_{DW}=\frac{T}{\frac{2\Pi}{\omega_{p}}}$.

The numerical scheme is a mixed Eulerian-Lagrangian method. Consequently, the mesh is subject to Lagrangian drift during the passage of steady,
steep wave groups, especially near the free-surface. For the transitory Class 3 wave groups defined in \citet{Song2002} and used during this present investigation,
the mesh did not deform significantly.

\subsection{Wave group generation }

\global\long\def\de#1#2{\dfrac{\partial#1}{\partial#2}}
 \global\long\def\De#1#2{\dfrac{d#1}{d#2}}

\global\long\def\dn#1{\dfrac{\partial#1}{\partial n}}

\global\long\def\dtdn#1{\dfrac{\partial^{2}#1}{\partial t\partial n}}

\global\long\def\ds#1{\dfrac{\partial#1}{\partial s}}

\global\long\def\dds#1{\dfrac{\partial^{2}#1}{\partial s{}^{2}}}

\global\long\def\dt#1{\dfrac{\partial#1}{\partial t}}

\global\long\def\Dt#1{\dfrac{d#1}{dt}}

\global\long\def\DDt#1{\dfrac{d^{2}#1}{dt{}^{2}}}

The WSIM wave-maker was chosen as a bottom-hinged flap paddle to generate
deep-water waves.

\begin{figure}
\psfrag{x_w}[][]  { $\mathbf{X_p\left(y\right)}$} %
\psfrag{r_g}[][]  { $\mathbf{r_g}$}%
\psfrag{\alpha}[][]  { $\mathbf{\alpha}$}%
\psfrag{x}[][]  { $\vec{\mathbf{x}}$}%
\psfrag{z}[][]  { $\vec{\mathbf{z}}$}%
\centering
\includegraphics[height=6cm]{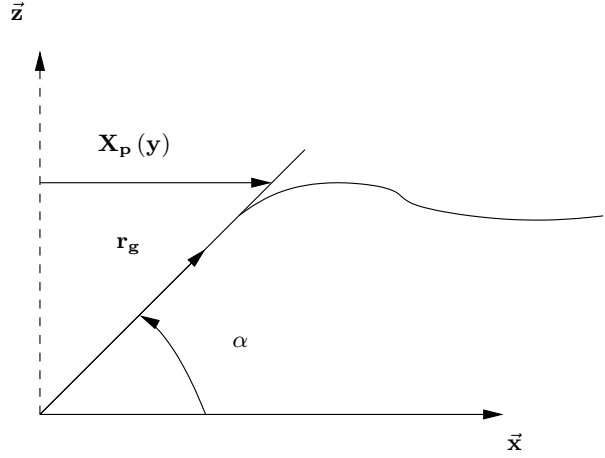}
\caption{2D Wavemaker}
\end{figure}

The wave-maker motion requires specification of the velocity potential
on the boundary as follows:

\begin{align}
\left.\dn{\phi}\right|_{bound.} & =r_{g}\dot{\alpha}\\
\left.\dtdn{\phi}\right|_{bound.} & =r_{g}\ddot{\alpha}+\dot{\alpha}\left[r_{g}\dds{\phi}-\ds{\phi}\right]
\end{align}
where $r_{g}$ is the distance of the centre of the rotation and $\alpha$ is the angle of rotation of the wave-maker.

To ensure smoothness of the boundary conditions and to avoid numerical instabilities at $t=0$ created by the moving elements, the paddle
oscillation motion $V_{wm(\theta,z)}$ is damped by a function $D\left(t\right)$ which changes smoothly from $0$ to $1$, as detailed
in \citet{Grilli1996} and \citet{Grilli1997}. Hence, the boundary condition can be written as follows:

\begin{align}
\left.\dn{\phi}\right|_{bound.} & =-V_{wm(\theta,z)}D(t)\\
\left.\dtdn{\phi}\right|_{bound.} & =-V_{wm(\theta,z)}\dot{D}(t)-\dt{V_{wm(\theta,z)}}D(t)
\end{align}

Because of finite tank-length considerations, we chose to investigate the chirped wave packet (Class 3) and investigated cases comprising five, seven and nine carrier
waves in the initial time wave packet.

\subsubsection{Class 3 wave groups}

Using the prescribed 'chirp' motion from \citet{Song2002}, the following expression defines Class 3 wave packets, $N$ being the number of waves
in the time series, $\omega_{p}$ is the paddle frequency and $\theta(t)=k\left[x_{w}-ct\right]-\theta_{0}$
is the phase. $C_{t2}$ is the chirp rate of the linear modulation:

\begin{align*}
Xp(t)= & \left(1+\tanh\left(\frac{4\omega_{p}t}{N\pi}\right)\right)\left(1-\tanh\left(\frac{4(\omega_{p}t-2N\pi)}{N\pi}\right)\right)\\
 & \sin\left(\omega_{p}\left(t-\frac{\omega_{p}C_{t2}t^{2}}{2}\right)+\Phi\left(X_{conv},Y_{conv}\right)\right).
\end{align*}

The paddle motion has an oscillating part moderated by a damping function.
To avoid discontinuities and assure a smooth numerical start in the
code, there is a negative shift in time of 5 seconds, to ensure that
any initial oscillations are of the order of the double precision
zero.

\subsubsection{3D converging wave group}

If 3D waves are to be produced by the snake paddle, the phase $\Phi$
becomes a function of $\left(X_{conv},Y_{conv}\right)$ which specify
the linear convergence point coordinates, as defined by \citet{Dalrymple1988,Dalrymple1989}:

\begin{eqnarray*}
\theta_{n} & = & \arctan\frac{y-Y_{conv}}{X_{conv}}\\
\Phi & = & k\cdot y\cdot \sin\theta_{n}+k\left(X_{conv}\cos\theta_{n}+Y_{conv}\sin\theta_{n}\right)
\end{eqnarray*}

\section{Post-processing}

\subsection{Determining near-surface quantities:}

A significant challenge within this investigation was determining
quantities close to the highly-curved free surface. Three complementary
solutions have been implemented. Evaluating the interior fields is
known to be a 'quasi-singular' problem, because it evaluates integration
with kernels of the type $\frac{1}{r^{\alpha}}$, $r$ being the distance
between the point where the field is calculated and a point on the
boundary and $\alpha$ is a positive exponent. The closer the evaluation
points are to the boundary, the larger these terms become and they
can possibly grow without bound.

\subsection{Interior fields}

\label{interiorfields}

The boundary element formulation of 3D potential problems implies:
\begin{align*}
a\left(\mathbf{x_{s}}\right)\phi\left(\mathbf{x_{s}}\right)= & \int_{S}\left(q\phi^{\star}-\phi q^{\star}\right)dS
\end{align*}

where $\mathbf{x_{s}}$ is the source point, $\phi\left(\mathbf{x_{s}}\right)$
is the potential; and $q\left(\mathbf{x}\right)=\dn{\phi}$ is the
derivative of $u$ along the unit outward normal $\mathbf{n}$ at
$\mathbf{x}$ on the boundary $S$. $S$ is the boundary of the region
$V$ of interest and boundary conditions concerning $\phi$ and $q$
are specified on $S$. $a\left(\mathbf{x_{s}}\right)=1$ when $\mathbf{x_{s}}\in V$
and $a\left(\mathbf{x_{s}}\right)=\frac{1}{2}$ when $\mathbf{x_{s}}\in S$
and $S$ is smooth at $\mathbf{x_{s}}$

The fundamental solutions $\phi^{\star}$ and $q^{\star}$ are defined
by: 
\begin{align*}
\phi^{\star}\left(\mathbf{x},\mathbf{x_{s}}\right)= & \dfrac{1}{4\pi r},\quad q^{\star}\left(\mathbf{x},\mathbf{x_{s}}\right)=-\dfrac{\left(\mathbf{r},\mathbf{n}\right)}{4\pi r^{3}}
\end{align*}
where $\mathbf{r}=\mathbf{x}-\mathbf{x_{s}}$ and $r=\left|\mathbf{r}\right|$.

The flux at a point $\mathbf{x_{s}}\in V$ is given by the potential
gradient: 
\begin{align}
\de{\phi}{\mathbf{x_{s}}}= & \int_{S}\left(q\de{\phi^{\star}}{\mathbf{x_{s}}}-\phi\de{q^{\star}}{\mathbf{x_{s}}}\right)dS\label{eq:Green}
\end{align}

where 
\begin{align*}
\de{\phi^{\star}}{\mathbf{x_{s}}}= & \dfrac{\mathbf{r}}{4\pi r^{3}},\quad\de{q^{\star}}{\mathbf{x_{s}}}=\dfrac{1}{4\pi}\left(\dfrac{\mathbf{n}}{r^{3}}-\dfrac{\left(\mathbf{r},\mathbf{n}\right)}{r^{5}}\right)
\end{align*}

Equations are discretised on the boundary $S$ by boundary elements
$S_{e}(e=1,N)$ defined by the interpolation function. The integral
kernels of these equations become nearly singular when the distance
$d=\left(\mathbf{r},\mathbf{n}\right)$ between $\mathbf{x_{s}}$
and $S_{e}$ is small compared to the size of $S_{e}$. In the following,
we will denote $S_{e}$ by $S$ for brevity.

Reconstructing the inner field values relies on the ability to correctly
estimate the sum \eqref{eq:Green} over the entire domain. Three complementary
methods are used depending on the distance $d$ between $\mathbf{x_{s}}$
and $S_{e}$.

When the distance $d$ is large enough (relative to the element size),
the integration on the individual element is carried out by a Riemann
quadrature on a classical Gauss-Lobatto point distribution.

When the distance $d$ becomes small, the classical quadrature does
not retain enough precision, and the \citet{Telles1987} method is
used. The Telles method consists of binary subdivisions of the integration
space to maximize precision where the singularity begins to manifest
itself. The precision of this method is acceptable at moderate distance
from the boundary, as described in \citet{Grilli1994}.

However, the Telles method becomes inefficient when the near-boundary
singularities are too strong, specifically adjacent to the highly
curved surface of steep waves. Consequently, a third method was developed
and implemented called Projection and Angular and Radial Transformation
(PART) (see \citet{Hayami1990}, \citet{Hayami1991}, \citet{Hayami1994},(\citet{Hayami2005}, \citet{Hayami2005a}). A change of variables is
used to calculate the integration on a closed element.

When applied to the curved quadrilateral element $S$, for the nearly-singular
integral over a curved boundary element:

\begin{align}
I & =\int_{S}\dfrac{f}{r^{\alpha}}dS\label{IntPART}
\end{align}

A sequence of steps is needed in order to effect a change of variables
and integrate: 
\begin{itemize}
\item 1. Source projection: Find the nearest point $\mathbf{\overline{x_{s}}}=\mathbf{x}\left(\overline{\eta_{1}},\overline{\eta_{2}}\right)$
on $S$ to $\mathbf{x_{s}}$ using the Newton-Raphson method. $\left(\overline{\eta_{1}},\overline{\eta_{2}}\right)$
are the local orthonormal coordinate system on the element $S$. 
\item 2. Determine the relative position of the source projection in the
local coordinate system. 
\item 3. Calculate the unit normal $\mathbf{n_{s}}$ to the curved element
$S$ at $\mathbf{\overline{x_{s}}}$ 
\item 4. Approximately project the curved element $S$ on the polygon $\overline{S}$
in the plane tangent to $S$ at $\mathbf{\overline{x_{s}}}$ 
\end{itemize}

\begin{figure}
\psfrag{A1}[][]  { $\mathbf{x_{1}}$}
\psfrag{A2}[][]  { $\mathbf{x_{2}}$}
\psfrag{A3}[][]  { $\mathbf{x_{3}}$}
\psfrag{A4}[][]  { $\mathbf{x_{4}}$}
\psfrag{Xs}[][]  { $\mathbf{x_{S}}$}
\psfrag{Ab1}[][] { $\mathbf{\overline{x_{1}}}$}
\psfrag{Ab2}[][] { $\mathbf{\overline{x_{2}}}$}
\psfrag{Ab3}[][] { $\mathbf{\overline{x_{3}}}$}
\psfrag{Ab4}[][] { $\mathbf{\overline{x_{4}}}$}
\psfrag{Xbs}[][] { $\mathbf{\overline{x_{S}}}$}
\psfrag{ns}[][]  { $\mathbf{n_{S}}$}
\psfrag{S}[][]  { $\mathbf{S}$}
\psfrag{Sb}[][]  { $\mathbf{\overline{S}}$}

\centering
\includegraphics[height=6cm]{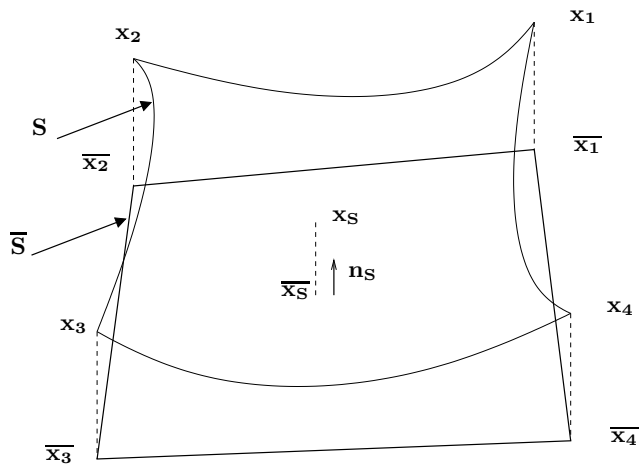}
\caption{\label{fig:PARTproj} Approximate projection.}
\end{figure}

\begin{itemize}
\item 5. Determine the geometry of the projected quadrilateral $\overline{S}$
and four triangular regions defined by joining $\mathbf{\overline{x_{s}}}$
to the four corners of $\overline{S}$, and then determine the linear
mapping matrices $L_{j}$. 
\item 6. Introduce polar coordinates $(\rho,\theta)$ in $\overline{S}$,
centered at $\mathbf{\overline{x_{s}}}$. Eq. (\ref{IntPART}) becomes: 
\end{itemize}
\begin{align}
I= & \int_{0}^{2\pi}d\theta\int_{0}^{\rho_{max}}\frac{f}{r^{\alpha}}J\rho d\rho\label{IntPART2}
\end{align}
where $J$ is the Jacobian of the mapping from Cartesian coordinates
on $\overline{S}$ to curvilinear coordinates on $S$. 
\begin{itemize}
\item 7. Apply radial variable transformation: $R(\rho)$ in order to weaken
the near-singularity due to $\frac{1}{r^{\alpha}}$, which is essentially
related to the radial variable $\rho$ only: 
\item 8. Apply the angular variable transformation $\left(\theta\right)$,
with $h_{j}$ being for each triangle $j$ the distance between $\mathbf{\overline{x_{s}}}$
and the foot of the perpendicular from $\mathbf{\overline{x_{s}}}$
and the edge of $\overline{S}$ and $\alpha_{j}$ the angle at $\mathbf{\overline{x_{s}}}$
of the triangle $j$: 
\end{itemize}
\begin{align}
t\left(\theta\right) & =\frac{h_{j}}{2}\log\left(\frac{1+\sin\left(\theta-\alpha_{j}\right)}{1-\sin\left(\theta-\alpha_{j}\right)}\right)
\end{align}

in order to weaken the angular near-singularity which arises from
$\rho_{max}(\theta)$ when $\mathbf{\overline{x_{s}}}$ is near the
edge of the polygon $\overline{S}$. This uses the fact that: 
\begin{align}
\De{\theta}t & =\dfrac{1}{\rho_{max}(\theta)}=\dfrac{\cos(\theta-\alpha_{j})}{h_{j}}
\end{align}

\begin{itemize}
\item 9. Use Gauss-Legendre's formula to perform numerical integration of
(\ref{IntPART2}) in the transformed variables $R$ and $t$ : 
\end{itemize}
\begin{align}
I & =\int_{t(0)}^{t(2\pi)}\frac{dt}{\rho_{max}(\theta)}\int_{R(0)}^{R\left(\rho_{max}(\theta)\right)}\frac{fJ\rho}{r^{\alpha}}\De{\rho}RdR
\end{align}

\subsection{Determination of physical quantities}

Local surface elevation and its corresponding depth-integrated potential
energy are obtained from the free surface data, extracted from the
data using the same 3rd order spline interpolation polynomials as
used in the BEM simulation (section \ref{discretisation}). Velocities
and the associated depth-integrated kinetic energy are calculated
using free surface data and the interior fields as seen above (section
\ref{interiorfields}).

The crest and trough locations are tracked using a two-stage detection
algorithm:

First, extrema and the zero-crossing positions are calculated directly
from the discretising spline polynomials. These positions are then
linked together between time steps to form characteristics of the
motion. All physical quantities are tracked along these characteristic
lines.

\section{Results and Discussion}

The numerical simulations undertaken are summarised in Table \ref{tab:Case}.

\begin{landscape}

    \begin{longtable}{L{2.5cm}C{0.9cm}C{1.5cm}C{1.7cm}C{1.9cm}C{1.2cm}C{1.2cm}C{1.2cm}L{7.3cm}}
    \textbf{Identifier} & \textbf{Class} & \textbf{Number of Waves} & \textbf{Approx. points per wave} & \textbf{Paddle amplitude A} & \textbf{$\text{X}_{\text{focus}}$/$\lambda_p$} & \textbf{$C_0$}& \textbf{$C_{p_0}$}& \textbf{Comments} \\
    \hline
    C3N5A0.05 & 3     & 5     & 8     & 0.05  & 2D       & 1.1014   & 1.2878    & {Small steepness case} \\
    C3N5A0.3 & 3     & 5     & 8     & 0.3   & 2D        & 1.1171   & 1.22878    & {Medium steepness case} \\
    C3N5A0.508 & 3     & 5     & 16    & 0.508 & 2D      & 1.2382   & 1.2124    & {} \\
    C3N5A0.511 & 3     & 5     & 16    & 0.511 & 2D      & 1.2499   & 1.2028    & {Marginal high resolution recurrence case} \\
    C3N5A0.513 & 3     & 5     & 8     & 0.513 & 2D      & 1.2024   & 1.2156    & {} \\
    C3N5A0.514 & 3     & 5     & 8/16  & 0.514 & 2D      & 1.2416   & 1.1978    & {Marginal high resolution breaking case} \\
    C3N5A0.516 & 3     & 5     & 8/16  & 0.516 & 2D      & 1.2380   & 1.2022    & {} \\
    C3N5A0.518 & 3     & 5     & 8/16  & 0.518 & 2D      & 1.2126   & 1.2023    &{Marginal medium resolution recurrence case} \\
    C3N5A0.519 & 3     & 5     & 8/16  & 0.519 & 2D      & 1.2275   & 1.1882    & {Marginal medium resolution breaking case} \\
    C3N5A0.53 & 3     & 5     & 8     & 0.53  & 2D       &          &           & {Breaking onset too close to paddle to obtain reliable values for $C_0$ and $C_{p_0}$} \\
    C3N5A0.56 & 3     & 5     & 8     & 0.56  & 2D       &          &           & {Breaking onset too close to paddle to obtain reliable values for $C_0$ and $C_{p_0}$} \\
    C3N5A0.32X10 & 3     & 5     & 8     & 0.32  & 10    & 1.3940   & 1.1901    & {} \\
    C3N5A0.33X10 & 3     & 5     & 8     & 0.33  & 10    & 1.2538   & 1.2290    & {Marginal recurrence case} \\
    C3N5A0.34X10 & 3     & 5     & 8     & 0.34  & 10    & 1.2472   & 1.2250    & {Marginal breaking case} \\
    C3N5A0.35X10 & 3     & 5     & 8     & 0.35  & 10    & 1.2568   & 1.2202    & {} \\
    C3N5A0.36X10 & 3     & 5     & 8     & 0.36  & 10    & 1.2735   & 1.2168    & {} \\
    C3N7A0.41 & 3     & 7     & 8     & 0.41  & 2D       & 1.2925   & 1.3141    & {} \\
    C3N7A0.42 & 3     & 7     & 8     & 0.42  & 2D       & 1.2974   & 1.3155    & {} \\
    C3N7A0.43 & 3     & 7     & 8     & 0.43  & 2D       & 1.2964   & 1.3233    & {} \\
    C3N7A0.44 & 3     & 7     & 8     & 0.44  & 2D       & 1.2981   & 1.3592    & {} \\
    C3N7A0.45 & 3     & 7     & 8     & 0.45  & 2D       & 1.3075   & 1.3082    & {} \\
    C3N7A0.46 & 3     & 7     & 8     & 0.46  & 2D       & 1.3029   & 1.3102    & {} \\
    C3N7A0.47 & 3     & 7     & 8     & 0.47  & 2D       & 1.3002   & 1.3505    & {Marginal recurrence case} \\
    C3N7A0.48 & 3     & 7     & 8     & 0.48  & 2D       & 1.3156   & 1.3085    & {Marginal breaking case} \\
    C3N7A0.49 & 3     & 7     & 8     & 0.49  & 2D       & 1.3430   & 1.3290    & {} \\
    C3N7A0.5 & 3     & 7     & 8     & 0.5   & 2D        & 1.2818   & 1.2593    & {} \\
    C3N9A0.42 & 3     & 9     & 8     & 0.42  & 2D       & 1.4651   & 1.4521    & {} \\
    C3N9A0.43 & 3     & 9     & 8     & 0.43  & 2D       & 1.4771   & 1.4942    & {} \\
    C3N9A0.44 & 3     & 9     & 8     & 0.44  & 2D       & 1.4836   & 1.5000    & {} \\
    C3N9A0.45 & 3     & 9     & 8     & 0.45  & 2D       & 1.4904   & 1.5054    & {} \\
    C3N9A0.46 & 3     & 9     & 8     & 0.46  & 2D       & 1.4938   & 1.4551    & {} \\
    C3N9A0.47 & 3     & 9     & 8     & 0.47  & 2D       & 1.5229   & 1.5265    & {Marginal recurrence case} \\
    C3N9A0.48 & 3     & 9     & 8     & 0.48  & 2D       & 1.6005   & 1.4276    & {Marginal breaking case} \\
    C3N9A0.49 & 3     & 9     & 8     & 0.49  & 2D       & 1.2659   & 1.4044    & {} \\
  \caption{Summary of computational experiments and their reference velocities $c_0$ and $c_{p_0}$. } \label{tab:Case}%
    \end{longtable}%

    \begin{longtable}{L{2.5cm}C{1.0cm}C{1.0cm}C{2.7cm}C{1.0cm}C{1.0cm}C{2.7cm}C{1.2cm}C{1.2cm}C{2.7cm}}
    \textbf{method} &$\omega_{0}$  & $\Delta \omega_{0}$ & \textbf{Interpolated} $\omega_{0}$ & $k_{0}$ & $\Delta k_{0}$  & \textbf{Interpolated} $k_{0}$  & $c_{0}=\frac{\omega_{0}}{k_{0}}$ & $\frac{\Delta c_{0}}{c_{0}}$ & \textbf{Interpolated} $c_{0}$  \\
    \hline
    local           & 5.3833 & 0.4141 & 5.2393  & 4.1493 & 0.5187 & 4.2357 & 1.2974  & 0.2020  & 1.2405 \\
    average         & 5.3833 & 0.4141 & 5.2238  & 4.1490 & 0.519  & 4.1793 & 1.2974  & 0.2020  & 1.2499 \\
    dispersion plot & 5.3833 & 0.4141 &         & 4.1493 & 0.5187 &        & 1.2974  & 0.2020  &        \\

\caption{\label{tab:c0}Determination of the reference value of the linear velocity $c_{0}$. This table shows the convergence of these values
$c_{0}$ using 3 different methods, local, on averaged data or using the whole dispersion plot, in the marginal recurrence C3N5A0.511 case. $\Delta\omega_{0}$
and $\Delta k_{0}$ are the spectral resolutions around the maximum.}
\end{longtable}    

\end{landscape}%

\subsection{Crest and trough slowdown}

\subsubsection{Local quantities}

\begin{figure}
\centering \includegraphics[width=0.49\textwidth]{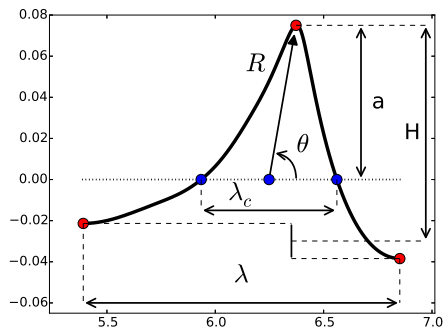}
\caption{\label{fig:realprofile} C3N5 wave profile with the definitions of the classical and local quantities as well as associated polar coordinates.}
\end{figure}

The local behaviour of the wave kinematics calls for precise definitions of wave geometry (Figure \ref{fig:realprofile}): amplitude $a$,
height $H$, wavelength $\lambda$ and crest half wavelength $\lambda_{c}$. The local crest steepness $S_{c}$ is defined as $S_{c}=a\frac{\pi}{\lambda_{c}}$.
An analogous definition exists for the local trough steepness $S_{t}$.



\begin{figure}
\centering
\includegraphics[width=0.49\textwidth,height=5cm]{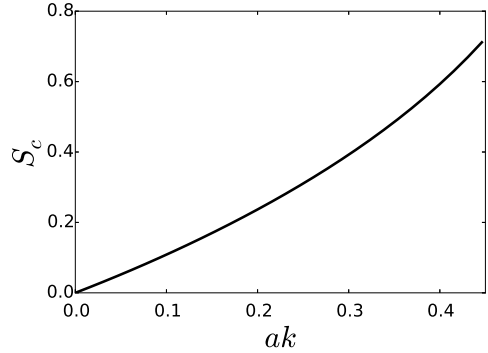}
\caption{\label{fig:stokesakvsSc}Relationship between local steepness $S_{c}$
and Stokes steepness $ak$ numerically computed for a $5^{th}$ order
Stokes wave.}
\end{figure}

Figure \ref{fig:stokesakvsSc} shows the relationship between the Stokes steepness $ak=H\frac{\pi}{\lambda}$ and the local crest steepness
$S_{c}$ calculated using a $5^{th}$order Stokes wavetrain \citep{Fenton1985}. It is noted that the classical Stokes limit of $ak=0.42$ becomes
a limiting steepness of $S_{c}=0.71$ when expressed in terms of the local crest steepness.

\subsubsection{Determination of the reference wave speed $c_{0}$}

For robust comparison, the reference wave speed $c_{0}$ is required and needs to be precisely and carefully determined. A spectral analysis
of both time and space wave signals is undertaken to be able to estimate the true reference wave quantities rather than those based on the
paddle signal.

\begin{figure}
\centering \includegraphics[width=0.49\textwidth,height=5cm]{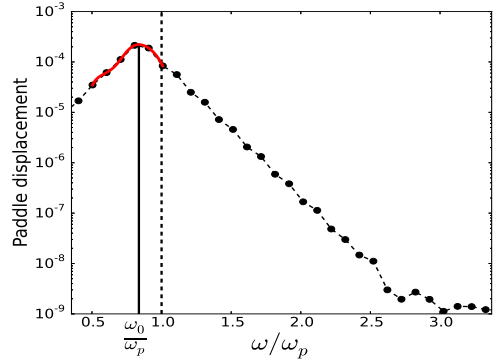}
\caption{\label{fig:PaddleStrokeFFT}Displacement spectrum for the Class 3 N5 paddle stroke signal. The damping component modulates the signal
and the spectral peak value $\omega_{0}$ (solid vertical line) differs from the generating wavenumber $\omega_{p}$ (dashed vertical line). The spectral peak was located accurately using 
a local spline (solid line).}
\end{figure}

Figure \ref{fig:PaddleStrokeFFT} shows the amplitude spectrum of a representative paddle signal. The characteristic generating frequency $\omega_{p}$
is systematically different from the power spectrum peak value $\omega_{0}$ (vertical black line) obtained by a FFT analysis. Consequently, each
investigated case has to be analysed in space and time to determine the reference power spectrum peak wavenumber $k_{0}$ and peak frequency $\omega_{0}$.

Several methods were used to determine the corresponding values of $c_{0}=\frac{\omega_{0}}{k_{0}}$ and $c_{p_{0}}=\frac{g}{\omega_{0}},$ and these show convergence towards identical values
for the same simulation case. The linear velocity $c_{0}$ has been estimated without making any hypothesis about the (deep-water) dispersion
relationship, using the peak of the frequency spectrum $\omega_{0}$ and the peak of the wave number spectrum $k_{0}$, giving the
linear celerity from the ratio $c_{0}$. The deep water dispersion relationship can be used to determine $c_{p_{0}}$, using the peak frequency $\omega_{0}$.

Table \ref{tab:c0} shows an example of three different methods to obtain $c_{0}$ using a local method, an average
method and a global method for the 2D marginal breaking C3N5A0.511 case.

\begin{figure}
\centering 
\includegraphics[width=0.95\textwidth]{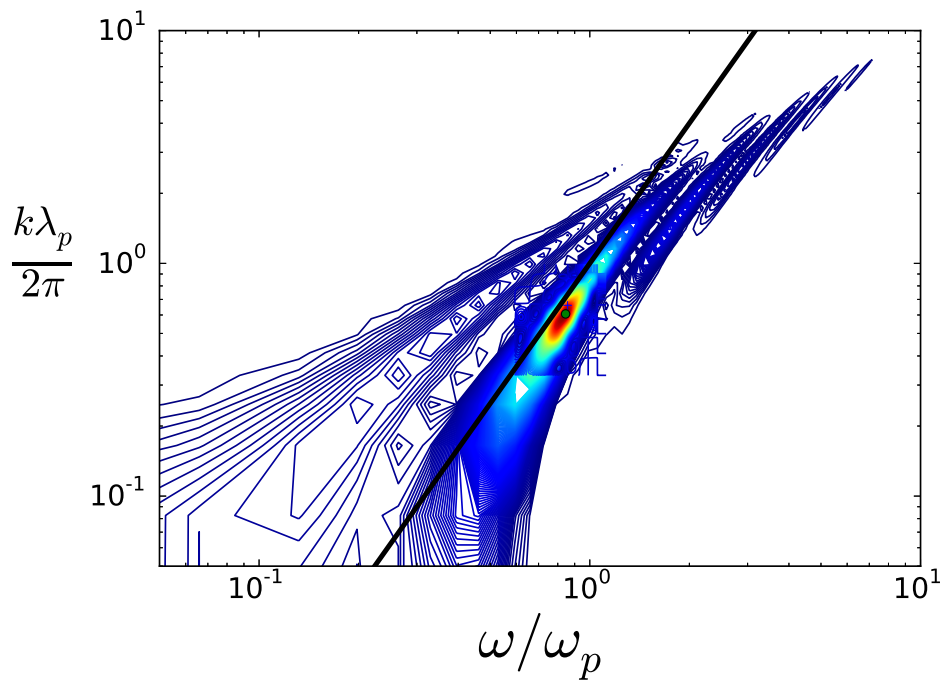}
\caption{\label{fig:Dispersion}Dispersion plot for the C3N5A0.511 marginal recurrence case. Colours represent the spectral energy density and the axes have
been normalised by the paddle frequency $\omega_{paddle}$ and the associated linear wavenumber $k_{paddle}=\omega_{paddle}^{2}/g$.
The coordinates of the peak in the graph are used to determine the reference linear velocity $c_{0}=\frac{\omega_{0}}{k_{0}}$. The solid
black line represent the Stokes linear deep water dispersion relationship}
\end{figure}

The local method value has been found by computing the spatial FFT at a random time and a FFT time series centered on the position of
the peak of the wave group at this given time. The average method is based on determining $\omega_{0}$ from the ensemble mean FFT of
the time series produced by 15 virtual height probes spaced equally, between two crest (or trough) maximum events. The peak wavenumber
$k_{0}$ was determined from the ensemble mean FFT of 512 surface profiles between two crest (or trough) maxima. The global method is a 2D FFT
of a space-time domain of the simulation to obtain the dispersion graph (figure \ref{fig:Dispersion}).

The values of $c_{0}$ found in Table \ref{tab:c0} show a very small sensitivity to the extent of the time and space domain used to determine
the reference velocity: $c_{0}$ remains almost constant. These frequency and wavenumber estimates are the raw values found by taking the maximum
values of the spectra at a given resolution. The length of the time series and the size of the domain do not provide sufficiently high resolution
in the spectral domain (indicated by $\Delta\omega_{0}$ and $\Delta k_{0}$), and a refined method was necessary. An interpolating polynomial is
constructed using the neighbouring points around the power spectral maximum and the new refined maximum is found analytically using the
polynomial coefficients. These new quantities give converged values of the reference velocities $c_{0}$ and $c_{p_{0}}$ that have been
tabulated for all the cases in the table \ref{tab:Case}.

Some additional remarks follow on the value of the reference velocity $c_{p_{0}}$ based on the deep water dispersion relation. Differences
between $c_{0}$ and $c_{p_{0}}$ have been found, which can be explained by the nature of the wave group. Each individual wave in the group
has a wavelength short enough to be considered in deep water in the simulation, but the wavelength of the wave-packet envelope is long
enough not to be considered in deep water. \citet{Longuet-Higgins1964} called this a transitional wave group. These differences are secondary
and do not affect the results.

In addition to the information about the spectral peak waves, the dispersion plot shows the degree of nonlinearity obtained with the
multiple higher frequency ``ripples''. These are the bound waves of the group, i.e. higher harmonics of the basic free waves constituting
the wave group. In general, the more harmonics that are present in the dispersion graph, the more nonlinear is the whole phenomenon.

\subsection{Space-time characteristics}

This study has been carried out using chirped Class 3 wave packets as defined above, but broader experimental and field studies (see
\citet{Banner2014}) confirm its generality.

\begin{figure}
\centering 
\includegraphics[width=0.99\textwidth]{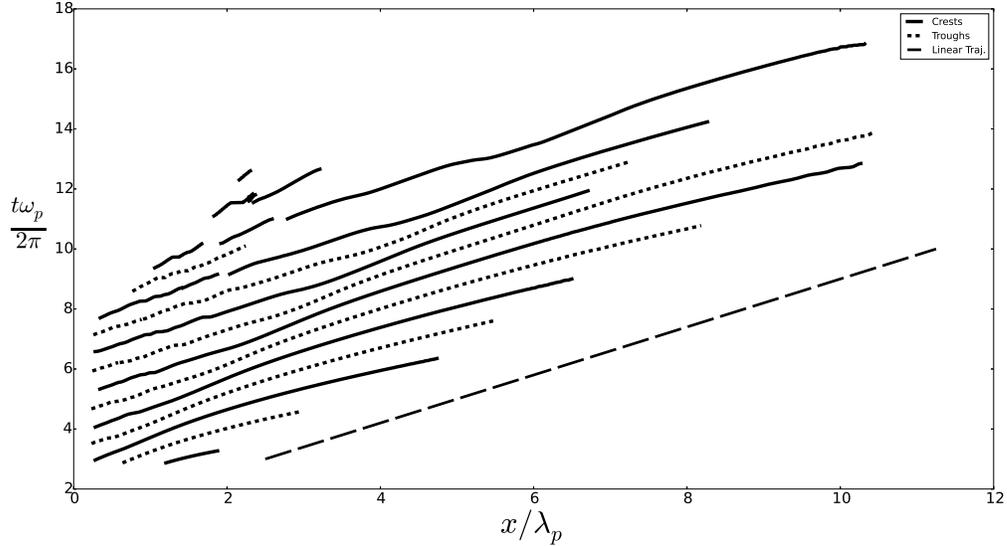}
\caption{\label{fig:spaghettiC3}Space-time plot of crests (full lines) and troughs (dashed lines) of the C3N5 marginal recurrence case. The long-dashed
line represents the reference linear velocity. Slowdown region occurs where the lines turn anticlockwise. In this plot, a systematic slowdown
and reacceleration of each crest and trough is visible.}
\end{figure}

Figure \ref{fig:spaghettiC3} represents the characteristics plot of crests and troughs along the centre slice of the 3D numerical wave
tank of a representative Class 3 wave packet. Crests, troughs and zero-crossings have been detected and followed in this coordinate
system, the abscissa is the non-dimensional position and the ordinate is the non-dimensional time. In this plot, the theoretical linear
trajectory is represented by a long-dash line as a reference, crest paths are the full lines and trough paths are represented by dashed
lines. Trajectories in space and time of each crest and trough are clearly seen as nonlinear, with local variations.

This plot exhibits systematic patterns for the crest and trough trajectories during the life cycle of each wave observed. The local slope of these
trajectories represents the velocity of the crest or trough. A trajectory path turning anticlockwise implies slowdown and a clockwise rotation
represents an acceleration. Each of the paths has a clear systematic variation in the slope and slowdown can be observed before a re-acceleration,
between $x/\lambda_{p}=1$ and $x/\lambda_{p}=5$. These slowdowns occur close to the crest and trough maxima of each wave inside the
wave packet.

\subsection{Slowdown of each wave, a generic feature.}

Slowdown phenomena have been observed and mentioned previously (\citet{Johannessen2001}, \citet{Johannessen2003}, \citet{Katsardi2011}), but have not been studied as a phase-resolved wave effect. The following plots address this gap by presenting the evolution of the crest (or trough) velocity as a function of the local steepness $S_{c}$. The
curves show the trajectory of the non-dimensional horizontal velocity of each carrier wave crest in the wavegroup, normalised by the theoretical
linear wavespeed $c_{0}$ defined above. Velocities are calculated using the first derivative of a 3rd-order spline fitting the space-time
trajectory. Results below are presented for crests only for illustration, and troughs are not shown because they behave in exactly the same
way.

\begin{figure}
\centering 
\includegraphics[width=0.75\textwidth]{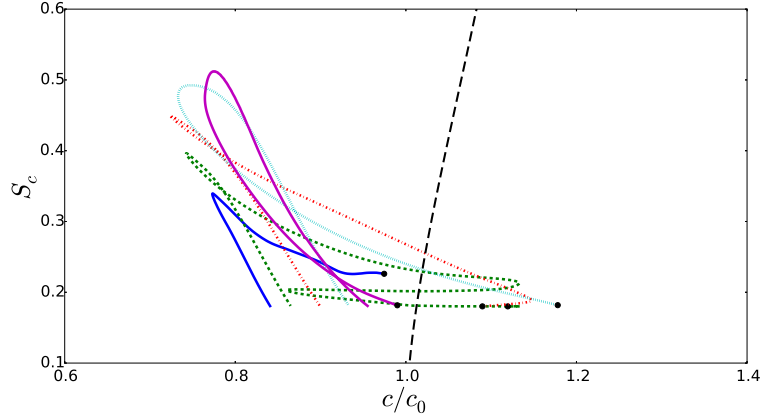}
\caption{\label{fig:C3Steepness-Vs-Celerity}Local steepness plotted against normalised velocity for a C3N5A0.511 marginal recurrent wavegroup. The dashed
line represents 5th order Stokes celerity for reference. Each of the lines represents an individual crest within the group which undertakes
an hysteresis cycle, showing a slowdown as the local steepness increases. The starting point is indicated by a black dot. Spikes seen in the
velocity trajectories are artefacts from the tracking algorithm and are explained below. }
\end{figure}

Figure \ref{fig:C3Steepness-Vs-Celerity} shows the evolution of the horizontal crest velocity of every crest of a Class 3, $N=5$ marginal
recurrent wave packet. The abscissa measures the non-dimensional velocity ratio $\frac{c}{c_{0}}$ and the ordinate shows the local steepness
$S_{c}$. Each individual crest life-cycle in the group is represented by a line trajectory. A black dot indicates where each trajectory starts.

Two important finding are represented in this plot showing the instantaneous crest velocity against the local steepness. The first conclusion,
which is one of the new findings of this study, is that contrary to inferences based on Stokes wavetrain theory, the crest velocity
is not strongly dependent on the steepness. Minimum crest velocity in the ensemble of waves shown in figure \ref{fig:C3Steepness-Vs-Celerity}
approaches the same value systematically, with a local steepness varying from $S_{c}=0.33$ to $S_{c}=0.52$. Previous knowledge of velocity
variation of crests is represented by the dashed line, which corresponds to $5^{th}$order Stokes crest celerity.

The second new result in this plot concerns the minimum crest velocity. In this marginal recurrent case, the velocities of each crest reduce
close to a minimum of $0.8$ of the linear reference velocity $c_{0}$ associated with this packet. The evolution of velocity is not linear
with the steepness, and loops can be seen in the curves describing the history of the crest velocity, which shows the existence of hysteresis
behaviour. It is noted that the minimum speed of this ensemble of crests seems to rise a few percent (from $\frac{c}{c_{0}}=0.75$ to
$0.78$ in this case) when the local steepness rises and the strong nonlinearities enter.

These findings are corroborated by \citet{Johannessen2001}, \citet{Johannessen2003}, \citet{Katsardi2011}. These authors show the existence of a slowdown
of 10\% in 2D waves and of 7\% in the case of 3D waves, with scatter. These slowdowns are quantified in respect to a reference linear velocity,
called $u_{l}$, which is not clearly defined. If a guess of the reference velocity estimation based on the peak of the frequency spectra
is made, then the normalising velocity becomes $c_{p_{0}}$ defined above. With this hypothesis, the minimum velocity for the dominant
crest in this graph goes from $\frac{c}{c_{0}}=0.78$ to $\mbox{\ensuremath{\frac{c}{c_{p_{0}}}=}}0.80$, resulting in a 20\% reduction in wave speed. A key question that arises
from this analysis is whether this effect is local. The above papers argue that the cause of the slowdown can be associated with the appearance
in the frequency wave spectrum of a peak around $1.2$ to $1.5$ of the main wave-packet peak, producing slower waves in accordance with the theoretical linear
free-wave dispersion relationship. However, the present study offers a complementary argument for the cause of the slowdown in section
\ref{section:leaning} below.

\begin{figure}
\centering
\includegraphics[width=0.99\textwidth]{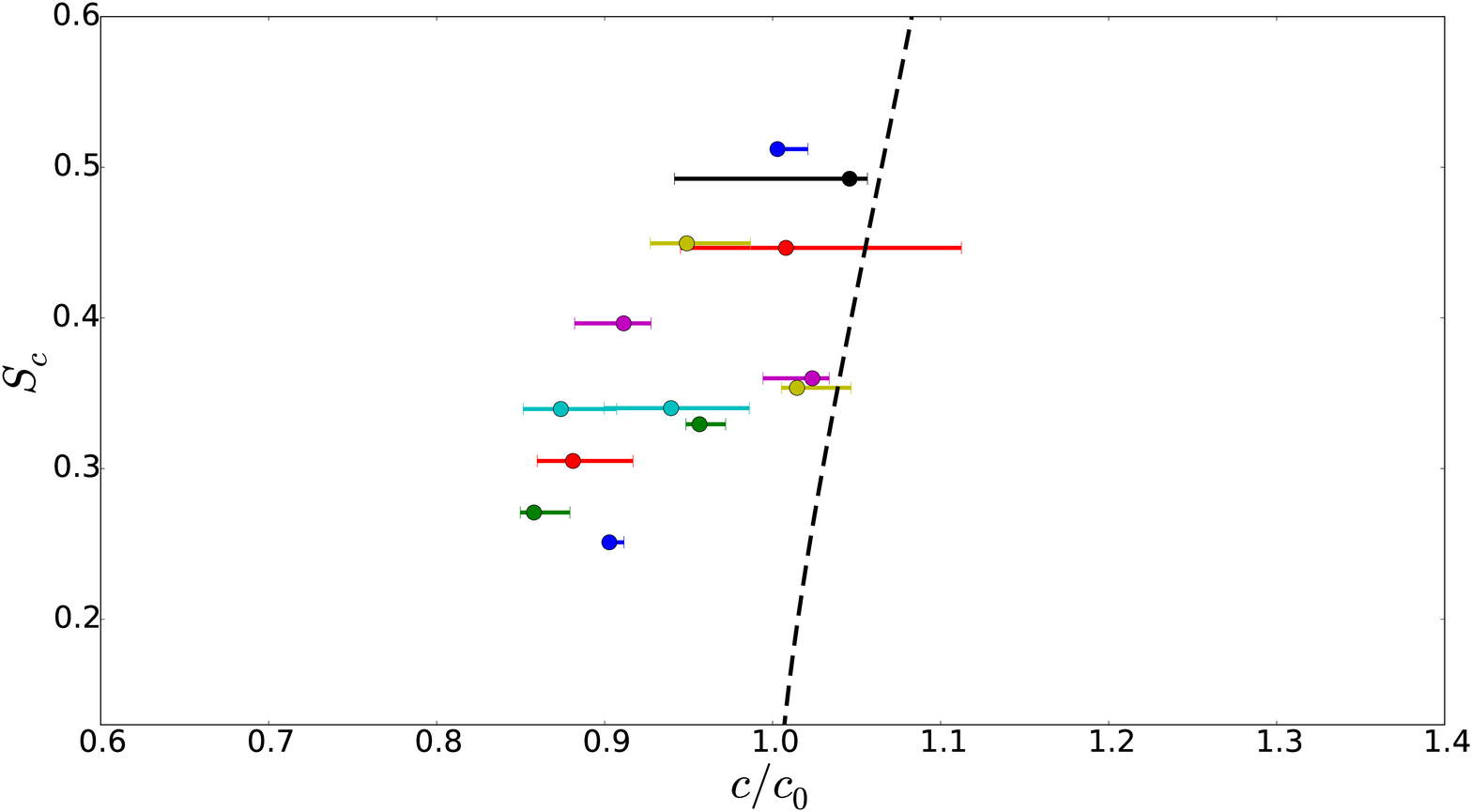}
\caption{\label{fig:Zero-crossings-middle}Zero-crossing mid-point velocity as a function of the maximum local steepness $S_{c}$ for each crest
for a range of 3D C3N5, 2D C3N5 and C3N7 cases, plotted against the crest velocity. The velocity at each crest maximum is represented by a 
dot, and the bars represent the extent of the variation of the velocity during each individual crest life-cycle.}
\end{figure}

The next figure addresses whether the slowdown is only a local crest and trough effect. Figure \ref{fig:Zero-crossings-middle} shows
the velocity of the mid-point between the two zero-crossings which adjoin (or span) each crest (or trough) studied. Each zero-crossing
pair has the steepness attribute of the crest they span. The abscissa is the non-dimensional velocity speed with respect to the reference
linear phase velocity $c_{0}$ .

The crest velocity at each crest maximum is represented by a dot, and the bars represent the extent of the variation of the
crest velocity during each individual crest life-cycle. The plot shows that the zero-crossing speed changes dynamically during
the wave evolution, and can undertake large relative speed variations of the order of roughly 20\%. Nonetheless, the speed values when the
crest reaches its maximum (dots) are close to the linear velocity, within the range $0.9 c_{0} - 1.0 c_{0}$. Consequently, the slowdown
is a localised crest effect, and does not affect the whole packet uniformly.

\subsection{Leaning, cause of the slowdown.\label{section:leaning}}

This section will show that the cause of the slowdown is geometric. We define the leaning as the left/right asymmetry :

\begin{align}
L & =-1+2\dfrac{X-X_{l}}{X_{r}-X_{l}}
\end{align}

where $X$ represents the crest (trough) position, $X_{l}$ is the position of the downward zero-crossing and $X_{r}$ is the position
of the upward zero-crossing. The leaning \emph{L} quantifies the asymmetry of the geometry of the triangle formed by the two zero-crossings spanning
the crest (or trough). This parameter varies from $-1$ when the backward-leaning face has a vertical asymptote to +1, when the forward-leaning face
is vertical.

Figure \ref{fig:realprofile} defines polar coordinates associated with a local crest (or trough). $\theta$ is the angle at the mid-point
(half-way point between the left and right zero-crossings) between the zero water level and the crest. $R$ is the radius, the distance
between the crest and the middle-point. Using polar decomposition of the velocity, the projection on the $x$ axis of the ortho-radial
component (the rotation part) is $C_{\theta x}=-R\Dt{\theta}\sin(\theta)$ and the projection on the $x$-axis of the radial component is $C_{Rx}=\Dt R\cos(\theta)$.

\begin{figure}
\centering
\includegraphics[width=0.95\textwidth]{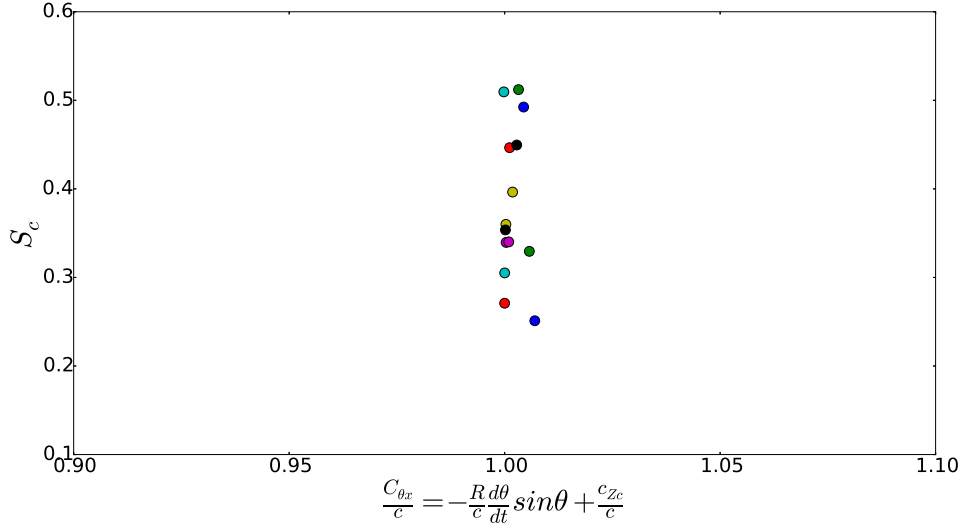}
\caption{Ratio of sum of $C_{\theta x}$ ($x$-axis projection of radial polar velocity) plus the zero-crossing speed plus $c$ (crest velocity
at maximum) against maximum steepness $S_c$ for an ensemble of crests for 3D C3N5, 2D C3N5 and C3N7 cases. Each dot represent an individual crest. This graph shows the crest slowdown is almost entirely included
in the radial relative velocity, implying that the rotation (leaning) is the cause of the slowdown.\label{fig:Celerity-and-leaning}}
\end{figure}

Figure \ref{fig:Celerity-and-leaning} shows the sum of the normalised frame of reference velocity $\frac{c_{Z_{c}}}{c}=\frac{1}{c}\Dt{X_{zc}}$
and the \emph{x}-projection of the normalised ortho-radial component of the relative velocity $\frac{C_{\theta x}}{c}=-\frac{R}{c}\Dt{\theta}\sin(\theta)$
at the crest maximum on the abscissa against the local steepness $S_{c}$ for C3 N5, C3N7, 2D and 3D wave groups. This plot shows a very strong
correlation between the rotation component and the crest velocity, and proves that the major part of the slowdown resides in the rotation
(hence the leaning) of crests about the mid-point between the zero-crossings. For all the waves in the group, each individual wave tends to lean
from forward to backward, reducing the crest velocity to below the expected linear crest velocity.

\begin{figure}
\centering
\includegraphics[width=0.99\textwidth]{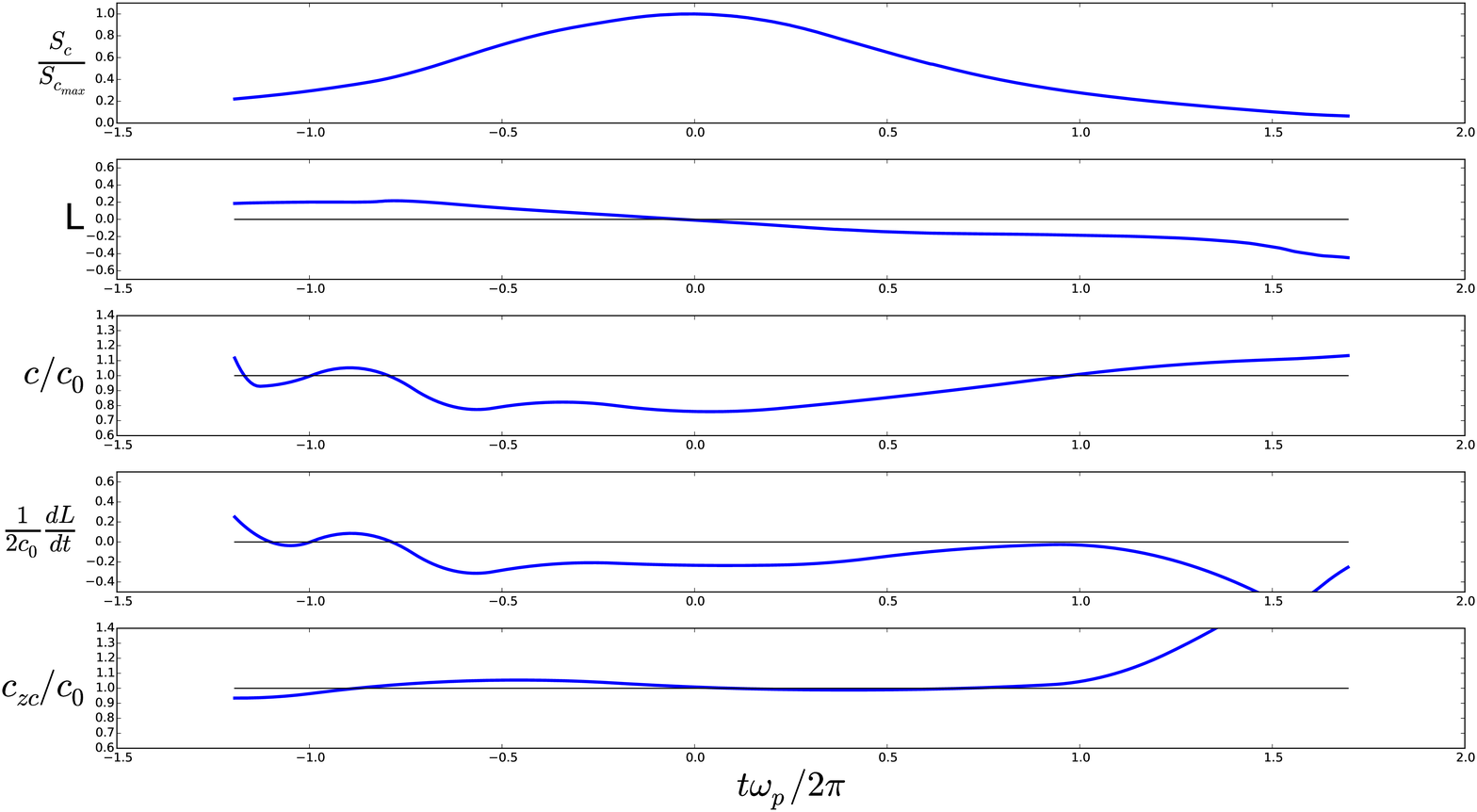}
\caption{Time evolution of the normalised wave properties tracking the dominant crest of a C3N50.511 marginally-recurrent wave group. The first panel is
the local crest steepness, the second panel is the leaning factor, the third panel is the mid-point zero-crossing velocity, the fourth
panel is the crest velocity and the bottom panel is the time derivative of the leaning. Note how well correlated are the quantities at crest
maximum (t=0): Leaning $L$ is 0, meaning the wave is symmetric, zero-crossing speed is close to 1.1, $c$ is around 0.8, and the time derivative
of $L$ indicates the symmetry is changing steadily backwards.}
\label{fig:Synchronisation} 
\end{figure}

As an example, figure \ref{fig:Synchronisation} summarises several key kinematic and geometric properties of the dominant crest of a
C3N5 wave group. In these graphs, there is a zero-reference time (along the abscissa) when the crest reaches its maximum elevation and the
local steepness is normalised to 1 at this crest maximum. The maximum local steepness for this wave is $S_{c}=0.512$.

This plot represents the entire history of a single carrier wave, with the first panel showing the normalised steepness and the second
panel showing the leaning parameter $L$. The third panel shows the average zero-crossing speed $\frac{c_{Zc}}{c_{0}}$ normalised by
the linear reference velocity, the fourth panel shows the crest velocity $\frac{c}{c_{0}}$ and the last panel is proportional to the leaning
velocity $\frac{g}{2c_{0}}\frac{dL}{dt}$. The graph reproduces the evolution history of the dominant crest travelling from the rear of
the wave group towards the front of the group.

The wave peaks at $t\omega_{p}/2\pi=0$. There is a strong correlation between the crest leaning backwards, which produces a negative time
derivative of $L$, and the crest slowdown, as seen between $t\omega_{p}/2\pi=-1$ and $t\omega_{p}/2\pi=1$. During the same time interval, the average
speed of the mid-point of the adjacent zero-crossings is almost constant, with a value close to the linear reference velocity. Some peaks appear
in the velocity and the time derivative of $L$, and are associated with a strong forward jump in $L$ at $t\omega_{p}/2\pi=-0.9$.
This is just an artefact of the crest-tracking process and is addressed in the section \eqref{sub:crestlet}.

One important finding is that $L=0$ always when the crest (or trough) reaches its maximum, implying that the wave crest (or trough) shape
is spatially symmetric at this instant. This leaning property could potentially be used to detect crest and trough maxima in a wave field.
These findings imply that the crest slowdown is a strong kinematic effect, easily understood via the composition of velocity law: the
frame of reference is the average position of the two adjacent zero-crossings travelling at roughly the linear phase velocity, and the local velocity
in this frame of reference is associated with the leaning forward-to-backward tendency of the crest, inducing a negative relative crest velocity.
The composition of the frame velocity and the backward leaning reduce the absolute crest velocity to below the originally-expected linear
phase velocity. Other fluctuations in the absolute velocity confirming this analysis can be seen just after $t\omega_{p}/2\pi=1$, where
a positive velocity of the frame of reference increases the absolute velocity above the linear level.

\begin{figure}
\centering
\includegraphics[width=0.99\textwidth]{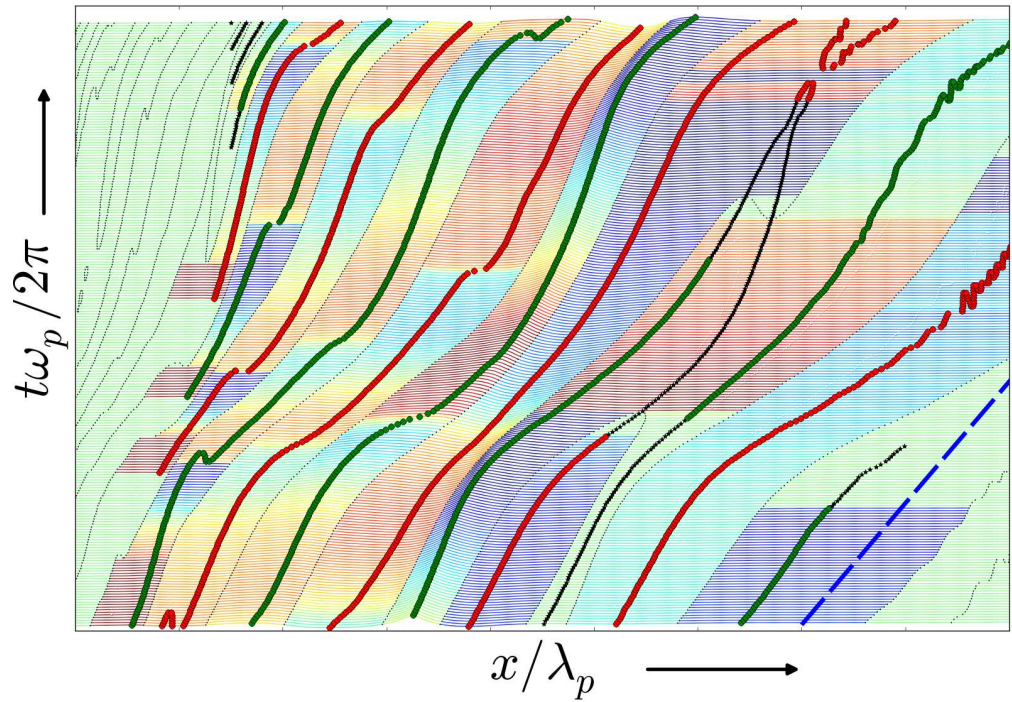}
\caption{Space-time plot of the evolution of the surface profile towards the C3N5 group maximum. Spatial profiles are stacked in time and colour-coded
to highlight the leaning dynamics. Green dots indicate crest positions, red dots indicate troughs, bold black stars are crests and troughs
below the threshold of detection and black dots show zero crossings. Leaning is colour-coded between zero-crossings: red (forward-leaning)
to blue (backward-leaning). The linear Stokes velocity $c_{0}$ is shown as a blue long-dashed line. Each crest and trough undertakes
a leaning cycle from leaning forward, peaking at symmetry and then leaning backwards. When the crests/troughs slow down (trajectories
turn anticlockwise), there is a clear change of symmetry when the extrema transition from leaning forward to leaning backwards. \label{fig:leaning-cycle} }
\end{figure}

Figure \ref{fig:leaning-cycle} presents multiple wave profiles stacked in order to form a space-time diagram of the typical evolution
along the centreline of a C3N5 maximum recurrent wave group. The axes have the non-dimensional position as the abscissa and non-dimensional
time as the ordinate. The wave profiles are colour-coded with the value of the leaning. This figure summarises the behaviour described
above: both crests and troughs undertake a systematic evolution throughout their life cycle. It can be generalised as a generic behaviour as
follows: a crest (or a trough) begins its life leaning forward (red shading), peaks (the geometry of the wave becomes symmetric (green
shading)) and decays leaning backwards (blue shading). Some abrupt jumps towards the front of the wave (seen as sudden changes of shading
colour) are also present and are explained in the following section \eqref{sub:crestlet}.

\subsection{Apparent scattering and detailed geometric analysis: crestlets or
riding waves?\label{sub:crestlet}}

\begin{figure}
\centering \includegraphics[width=0.99\textwidth]{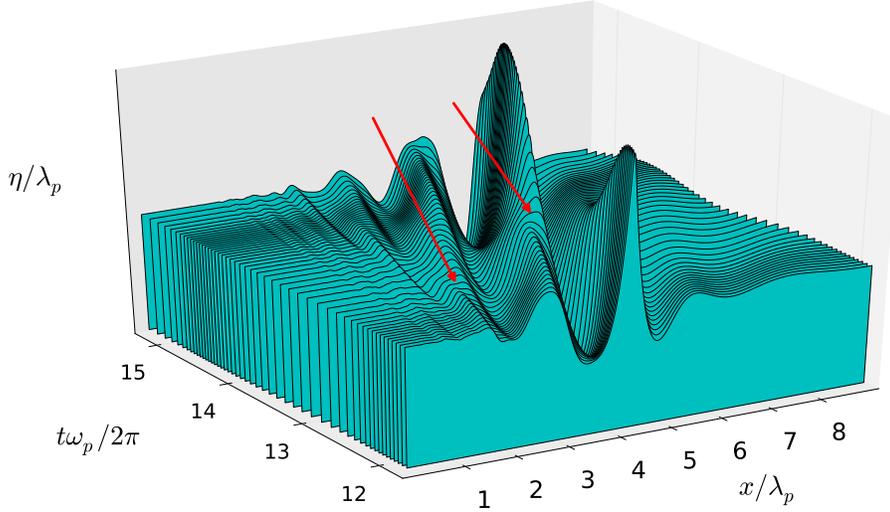}
\caption{ Space-time plot of the centreline of the 2D C3N5 marginal recurrent case. 2D profiles are stacked in the time direction. Spikes found
in celerity and leaning values are explained by the crest-trough tracking algorithm. A crest/trough has an evolution characterised by a succession
of 'crestlets' which appear on the forward face and propagate backwards, and will become dominant eventually. Each time the forward crestlet
becomes dominant, the detected crest position suddenly jumps forward creating these spikes, as highlighted by the red arrows.\label{fig:2D-1/2-plot} }
\end{figure}

This section explains the apparent velocity peaks found in the velocity plots (see Fig \ref{fig:Synchronisation} around $t=-0.8$, also apparent
in Fig \ref{fig:C3Steepness-Vs-Celerity} and Fig \ref{fig:leaning-cycle}). Figure \ref{fig:2D-1/2-plot} is a space-time
plot of the same wave packet shown in Figure \ref{fig:leaning-cycle}. The abscissa represents the position, the ordinate represents the time
coordinate and the depth represents the non-dimensional free surface elevation of a slice along the centreline of the numerical
wave tank. The plot is a time sequence of wave profiles.

In this section we explain the limitation of the crest-tracking algorithm and the apparent velocity spikes. The red arrows in Figure 14 identify
where the phenomenon occurs. The idea of a crest or a trough being a single maximum or minimum between two zero-crossings (which is an
arbitrary reference) refers to a Stokes wave. Natural unsteady waves sometimes have multiple crests and troughs between the same two zero-crossings,
which evolve dynamically. This effect can be seen in \citet{Johannessen2010} in their observational surface elevation plots. Also, \citet{Bateman2012}
describes a numerical model results where crestlets and leaning in the plots can be seen in their figures 8 and 9.

During the growth of a crest, there is an interchange of these small crests or troughs, the first one evolving and then the next one appearing
from the front of the packet to become temporarily the dominant crest, and then the next one appears. These crests are initiated, in the
relative frame of reference of the wave group, at the front of the group and then travel rearwards. During this crest exchange, which
is rapid (roughly a tenth of a time unit over a tenth of the reference length), the tracking system jumps from the decaying crest to the
next growing crest, sometimes going through a 'flattened' area of the crest. This creates a discontinuity in the detected crest position,
as it is now located further towards the front of the group. This produces the apparent high velocity spike. It should be noted that
between each of these peaks, the celerity of these "crestlets" decreases, back to $0.8c_{0}$.

The cause of these leaning effects is not yet clear, but possible insight is provided by \citet{Bettini1983} who a discuss the radiative
tail of a soliton. The radiative tail is the train of trailing waves produced by a propagating wave group. In our application, these forward
jumps of the crests (or troughs) align on the characteristic plot (Figure \ref{fig:spaghettiC3}). The average slope gives a velocity
one half of the linear velocity, this being the definition of the radiative tail in Bettini's article. These crestlets (or troughlets)
seem to propagate backwards. This effect is relative because their velocity is one half of the velocity of the wave crests (or troughs),
which causes the illusion. This energy leakage creates the radiative tail.

\section{Subsurface kinematics}

\subsection{Velocity profiles}

\begin{figure}
\centering
\includegraphics[width=0.7\textwidth]{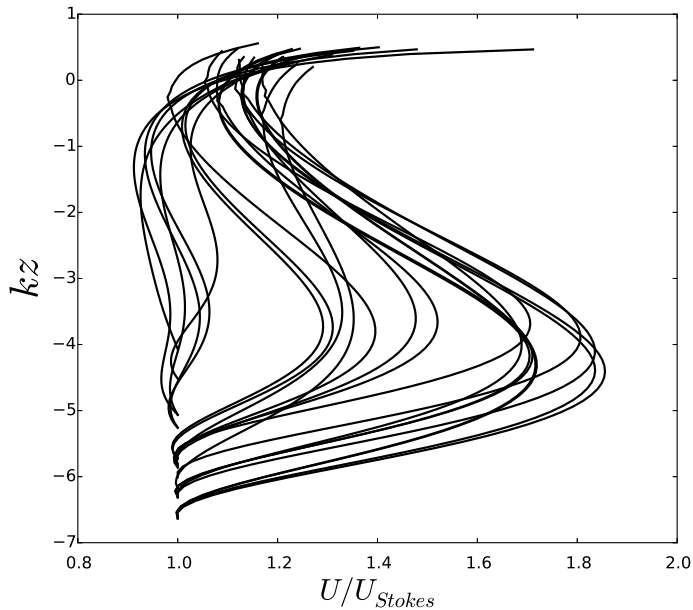}
\caption{Ratio of computed velocity profiles of a catalogue of cases beneath wave crests to the equivalent
$5^{th}$ order Stokes velocities, as a function of normalised depth.
The computed values have been corrected for the Eulerian drift \label{fig:UonUstokes}}
\end{figure}

This section provides further details on the subsurface motion by investigating the velocity profiles. Figure \ref{fig:UonUstokes}
shows the ratio of the velocity profiles computed from the NWT simulation and the 5th order Stokes velocity profiles for different cases. Shown
here are results below the local crest maxima of 2D C3N5 and C3N9 breaking and non-breaking (low steepness to marginal recurrence) ,
as well as some 3D cases. The Stokes waves used for comparison were fitted according to height $H$ and using the neighbouring troughs
to define the wavelength. Stokes velocity profiles have been corrected for the Eulerian current by matching velocities at the bottom of the tank.

Different zones can be identified. The first zone, around $kz=-4$, shows a relative ratio which can grow up to $1.8$. While this value
appears substantial, the actual velocity differences are not very large because the absolute values of the velocities are small, due
to the quasi-exponential attenuation of the velocity with depth.

The second zone, above $kz=0$, is more important as it is close to the free surface. Here the ratio varies between $1.2$ for maximum
recurrence cases and $1.7$ when the wave breaks. This change is significant and highlights an important difference between the predicted Stokes
orbital velocity and the actual orbital velocity computed by the simulation. The consequence can be important when estimating the impact of a wave
on a structure where the drag is proportional to the square of the velocity. The Stokes fit systematically underestimates the particle
crest velocity by at least 20\% for very steep waves, suggesting a near-halving of the impulse loading.

\section{Energy partitioning}

One of the main properties commonly assumed for waves is equipartitioning of the mean kinetic energy (KE) and potential energy (PE). With this
hypothesis, meaurements can be made that only measure wave heights (hence PE), from which the kinetic energy is deduced. This equipartitioning
is theoretically true for steady linear wave trains, but does it hold for a nonlinear unsteady wave group? This section investigates the
energy partitioning in a global sense, first analysing the whole wave group and then studying locally what occurs at the scale of the individual
crest and troughs.

\subsection{Global partitioning}

\begin{figure}
\centering
\includegraphics[width=0.7\textwidth]{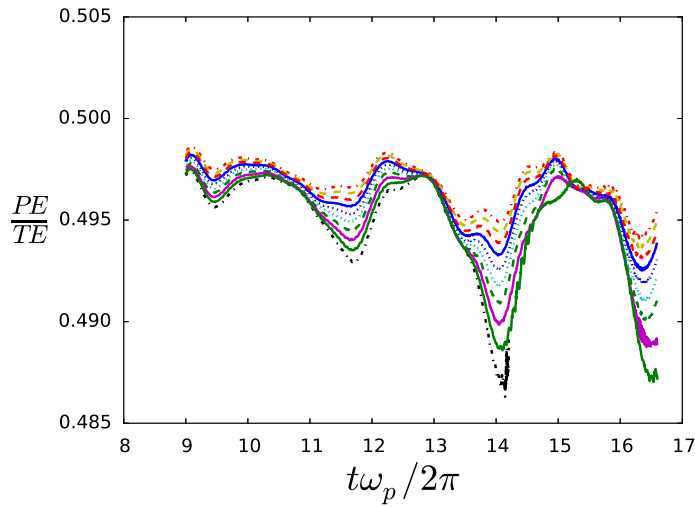}
\caption{\label{fig:totalPE} Depth-integrated energy partitioning averaged over the wave group for multiple 2D C3N7 cases. The evolution of breaking
cases is shown by the bottom dot-dashed line ending at $t=14.2$, and the second (solid) line from the bottom  and the third (solid) line from the bottom, both terminating at $t=16.3$. The ratio of
potential energy to total energy of these unsteady, nonlinear wave groups shows that linear equipartitioning is almost conserved at this
scale. Small variations are evident, with breaking groups tending to have slightly more kinetic energy as the ratio in these cases decreases to $0.49$}
\end{figure}

Having proven in the simulation validation section that energy conservation was respected, further investigation was undertaken concerning the
energy partitioning for the whole wave group. Figure \eqref{fig:totalPE} shows the depth-integrated energy partition $\frac{PE}{TE}$ for the
whole wave group with respect to time, for the C3N7 breaking and non-breaking cases. Periods when the whole group was not in the integration domain
have been blanked out. This figure shows that global equipartitioning holds with a small excess in the kinetic energy at the crest maximum.
The ratio oscillates globally close to $\frac{PE}{TE}=0.5$, showing that the total energy (TE) is almost equally divided between potential
energy (PE) and kinetic energy (KE). Breaking cases have the maximum deviation from equipartitioning with $\simeq 49 \%$ of the
energy residing in the PE. The energy flux necessary to produce the jet formed as the breaking event initiates has its source in the kinetic
energy.

\subsection{Local energy partitioning at crests and troughs}

\begin{figure}
\centering
\includegraphics[scale=0.55]{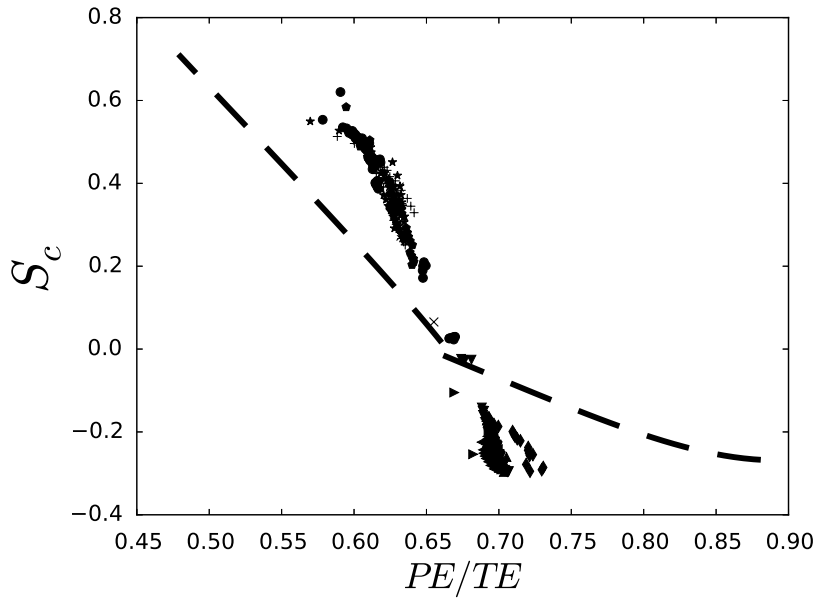}
\caption{\label{fig:PE/TE-local} Depth-integrated energy partitioning for the dominant crests (C3N5 are circles, C3N7 are stars, C3N9 are pluses, C3N10 are x-crosses and 3D C3N5 are pentagons) and troughs (C3N5 are inverted triangles, C3N7 are triangles, C3N9 are left-pointing triangles, C3N10 are right-pointing triangles and 3D C3N5 are diamonds) at local group
maximum for a catalogue of cases. The dashed lines represent this ratio for a 5th order Stokes wave. A clear deviation from linear theory
energy partitioning is seen, with 2D troughs clustering around a ratio of local depth-integrated potential energy to local depth-integrated
total energy of $0.7$, and 3D troughs around $0.71$. Also, 2D and 3D crests cluster from $0.6$ to $0.64$. This ratio drops below $0.6$ at the initiation
of breaking crests, signalling an excess of KE. Small amplitude waves have the same ratio of $0.67$ for crests and troughs.}
\end{figure}

The above section showed how the depth-integrated energy partitioning still holds for the whole unsteady wave group. We now examine the
local behaviour of the depth-integrated energy partitioning at both crest and trough maxima. Figure \ref{fig:PE/TE-local} shows the
energy partition $\frac{PE}{TE}$ plotted against local steepness $S_{c}$ for 2D and 3D breaking for the maximum recurrent wave packet
cases C3N5, C3N7, C3N9 and C3N10. The dashed lines represent this ratio for a 5th order Stokes wave. Triangles represent the values
for the troughs and circles show the values for the crests. Remarkably, our results show that equipartitioning breaks down locally at crest
and trough maxima and the $\frac{PE}{TE}$ ratio stays almost constant, $0.6$ to $0.62$ for the crests and 0.7 for the troughs. Locally, the Stokes
wave does not conform to the equipartitioning, but behaves differently from the unsteady group waves. The $\frac{PE}{TE}$ ratio for the
unsteady wave groups matches the Stokes wave for very small steepness and is close to $0.67$, but diverges from this level at higher steepness.
The notable difference between the marginal recurrent case and the breaking case is the breaking crest acquires proportionally more kinetic
energy with the partition ratio falling below $0.6$. It is again evident that the source of the breaking jet resides in the kinetic energy,
while the wave height (hence the potential energy) remains almost constant.

\section{Conclusions and recommendations}

This computational study of highly nonlinear gravity water waves unsteady wave groups has revealed new insights into the behaviour of kinematic
and geometric properties, as well as the associated energetics. In particular, our findings on generic crest slowdown obtained from these
numerical chirped wave group simulations have been confirmed observationally for chirped as well as for bimodal wave groups in our laboratory wave
basin experiments, as well as for deep water wave groups measured in an ocean tower experiment (\citet{Banner2014}).

When the theoretical constraint of steadiness is relaxed, additional asymmetric degrees of freedom are observed in the wave shape. In particular,
the kinematics reveal a forward-to-backward leaning cycle for each individual crest and trough inside the wave group, resulting in a
local crest or trough slowdown of around 20\% of the reference linear velocity estimated from a dispersion plot. This is a geometric effect,
as the velocity of the zero-crossings remains close to the linear phase velocity and the observed relative velocity variations can be
explained by the geometric changes in the waveform. Velocities profiles under the crest show a large difference between the Stokes-fit velocity
profile and the computed velocity profile. Computed orbital velocities under the crest can be significantly higher than for the Stokes wave
of corresponding height, and the ratio between them varies between 1.2 and 1.8 for the steepest cases. These values should be taken in
account when designing structures close to the surface of the ocean where the Stokes values are often assumed. The associated energetics
have been carefully monitored and support, within 1\%, equipartitioning of the depth-integrated energy on the scale of the wave group, although
at breaking onset, a slight excess of kinetic energy occurs. This depth-integrated energy equipartitioning breaks down when investigated
locally at crest and trough maxima during the evolution of the wave group. These show an energy partitioning of around 69\% for the potential
energy at trough maxima and a partitioning range from 60\% to 64\% of the potential energy at crest maxima. The crest energy
partitioning at breaking drops below 60\% for the potential energy, indicating the initiation of the breaking jet.

Overall, these findings contribute new insights into how actual ocean wave groups propagate when the academic constraint of steadiness is
relaxed. Aside from its intrinsic interest, the generic crest slowdown explains the reduced speed of initiation of breaker fronts that has
been reported occasionally in the literature and is central to assimilating whitecap data accurately into spectral sea-state forecast models.
Also, parameterisations of air-sea fluxes of momentum and energy, which depend on the square and cube of the sea-surface fluid velocity,
may also be modified appreciably, as may the impact loading on offshore structures.

\section{Acknowledgements}
Funding for this investigation was provided by the Australian Research Council under Discovery Project DP120101701. Also, FD acknowledges partial support by the European Research Council (ERC) under the research project ERC-2011-AdG 290562-MULTIWAVE and Science Foundation Ireland under grant number SFI/12/ERC/E2227.

\bibliographystyle{jfm}
\bibliography{biblio-full}

\begin{thebibliography}{71}
\expandafter\ifx\csname natexlab\endcsname\relax\def\natexlab#1{#1}\fi

\bibitem[Baker {\em et~al.\/}(1982)Baker, Meiron \& Orszag]{Baker1982}
{\sc Baker, Gregory~R., Meiron, Daniel~I. \& Orszag, Steven~A.} 1982
  Generalized vortex methods for free-surface flow problems. {\em Journal of
  Fluid Mechanics\/} {\bf 123}, 477--501.

\bibitem[Baldock {\em et~al.\/}(1996)Baldock, Swan \& Taylor]{Baldock1996}
{\sc Baldock, T.~E., Swan, C. \& Taylor, P.~H.} 1996 A laboratory study of
  nonlinear surface waves on water. {\em Philosophical Transactions:
  Mathematical, Physical and Engineering Sciences\/} {\bf 354}~(1707), pp.
  649--676.

\bibitem[Banner {\em et~al.\/}(2014)Banner, Barthelemy, Fedele, Allis,
  Benetazzo, Dias \& Peirson]{Banner2014}
{\sc Banner, M.\, L., Barthelemy, X., Fedele, F., Allis, M., Benetazzo, A.,
  Dias, F. \& Peirson, W.\, L.} 2014 Linking reduced breaking crest speeds to
  unsteady nonlinear water wave group behavior. {\em Phys. Rev. Lett.\/} {\bf
  112}, 114502.

\bibitem[Banner \& Peirson(2007)]{Banner2007}
{\sc Banner, Michael~L. \& Peirson, William~L.} 2007 Wave breaking onset and
  strength for two-dimensional deep-water wave groups. {\em Journal of Fluid
  Mechanics\/} {\bf 585}, 93--115.

\bibitem[Bateman {\em et~al.\/}(2012)Bateman, Katsardi \& Swan]{Bateman2012}
{\sc Bateman, W.J.D., Katsardi, V. \& Swan, C.} 2012 Extreme ocean waves. part
  i. the practical application of fully nonlinear wave modelling. {\em Applied
  Ocean Research\/} {\bf 34}~(0), 209 -- 224.

\bibitem[Bateman {\em et~al.\/}(2001)Bateman, Swan \& Taylor]{Bateman2001}
{\sc Bateman, W.J.D., Swan, C. \& Taylor, P.H.} 2001 On the efficient numerical
  simulation of directionally spread surface water waves. {\em Journal of
  Computational Physics\/} {\bf 174}~(1), 277 -- 305.

\bibitem[Benjamin \& Feir(1967)]{Benjamin1967}
{\sc Benjamin, T.~Brooke \& Feir, J.~E.} 1967 The disintegration of wave trains
  on deep water part 1. theory. {\em Journal of Fluid Mechanics\/} {\bf 27},
  417--430.

\bibitem[Bettini {\em et~al.\/}(1983)Bettini, Minelli \& Pascoli]{Bettini1983}
{\sc Bettini, Alessandro, Minelli, Tullio~A. \& Pascoli, Donatella} 1983
  Solitons in undergraduate laboratory. {\em American Journal of Physics\/}
  {\bf 51}~(11), 977--984.

\bibitem[Bey{\'a} {\em et~al.\/}(2012)Bey{\'a}, Peirson \& Banner]{Beya2012}
{\sc Bey{\'a}, J.F., Peirson, W.L. \& Banner, M.L.} 2012 Turbulence beneath
  finite amplitude water waves. {\em Experiments in Fluids\/} {\bf 52},
  1319--1330.

\bibitem[Clamond \& Grue(2001)]{Clamond2001}
{\sc Clamond, Didier \& Grue, John} 2001 A fast method for fully nonlinear
  water-wave computations. {\em Journal of Fluid Mechanics\/} {\bf 447},
  337--355.

\bibitem[Craig \& Sulem(1993)]{Craig1993}
{\sc Craig, W. \& Sulem, C.} 1993 Numerical simulation of gravity waves. {\em
  Journal of Computational Physics\/} {\bf 108}~(1), 73 -- 83.

\bibitem[Dalrymple(1989)]{Dalrymple1989}
{\sc Dalrymple, R.~A.} 1989 Directional wavemaker theory with sidewall
  reflection. {\em Journal of Hydraulic Research\/} {\bf 27}~(1), 23--34.

\bibitem[Dalrymple \& Kirby(1988)]{Dalrymple1988}
{\sc Dalrymple, Robert~A. \& Kirby, James~T.} 1988 Models for very wide-angle
  water waves and wave diffraction. {\em Journal of Fluid Mechanics Digital
  Archive\/} {\bf 192}~(-1), 33--50.

\bibitem[Dias {\em et~al.\/}(2015)Dias, Brennan, {Ponce de Leon}, Clancy \&
  Dudley]{Dias2015}
{\sc Dias, F., Brennan, J., {Ponce de Leon}, S., Clancy, C. \& Dudley, J.} 2015
  Local analysis of wave fields produced from hindcasted rogue wave sea states.
  In {\em Proceedings of the ASME 2015 34th International Conference on Ocean,
  Offshore and Arctic Engineering OMAE2015\/}.

\bibitem[Dommermuth \& Yue(1987)]{Dommermuth1987}
{\sc Dommermuth, Douglas~G. \& Yue, Dick K.~P.} 1987 A high-order spectral
  method for the study of nonlinear gravity waves. {\em Journal of Fluid
  Mechanics\/} {\bf 184}, 267--288.

\bibitem[Ducrozet {\em et~al.\/}(2012)Ducrozet, Bonnefoy, {Le Touze} \&
  Ferrant]{Ducrozet2011}
{\sc Ducrozet, Guillaume, Bonnefoy, Felicien, {Le Touze}, David \& Ferrant,
  Pierre} 2012 A modified high-order spectral method for wavemaker modeling in
  a numerical wave tank. {\em European Journal of Mechanics - B/Fluids\/} {\bf
  34}, 19--34.

\bibitem[Fedele(2014)]{Fedele2014}
{\sc Fedele, Francesco} 2014 Geometric phases of water waves. {\em EPL
  (Europhysics Letters)\/} {\bf 107}~(6), 69001.

\bibitem[Fenton(1985)]{Fenton1985}
{\sc Fenton, J.} 1985 A fifth-order stokes theory for steady waves. {\em
  Journal of Waterway, Port, Coastal, and Ocean Engineering\/} {\bf 111}~(2),
  216--234.

\bibitem[Fochesato(2004)]{Fochesato2004}
{\sc Fochesato, Christophe} 2004 Mod{\`e}les num{\'e}riques pour les vagues et
  les ondes internes. PhD thesis, CMLA / Ecole Normale Supérieure de CACHAN.

\bibitem[Fochesato \& Dias(2006)]{Fochesato2006}
{\sc Fochesato, Christophe \& Dias, Fr\'ed\'eric} 2006 A fast method for
  nonlinear three-dimensional free-surface waves. {\em Proceedings of the Royal
  Society A: Mathematical, Physical and Engineering Science\/} {\bf
  462}~(2073), 2715--2735.

\bibitem[Fochesato {\em et~al.\/}(2007)Fochesato, Grilli \&
  Dias]{Fochesato2007}
{\sc Fochesato, Christophe, Grilli, St{\'e}phan \& Dias, Fr{\'e}d{\'e}ric} 2007
  Numerical modeling of extreme rogue waves generated by directional energy
  focusing. {\em Wave Motion\/} {\bf 44}~(5), 395 -- 416.

\bibitem[Fructus {\em et~al.\/}(2005)Fructus, Clamond, Grue \&
  Kristiansen]{Fructus2005}
{\sc Fructus, Dorian, Clamond, Didier, Grue, John \& Kristiansen, Oyvind} 2005
  An efficient model for three-dimensional surface wave simulations: Part i:
  Free space problems. {\em Journal of Computational Physics\/} {\bf 205}~(2),
  665 -- 685.

\bibitem[Gemmrich {\em et~al.\/}(2008)Gemmrich, Banner \&
  Garrett]{Gemmrich2008}
{\sc Gemmrich, Johannes~R., Banner, Michael~L. \& Garrett, Chris} 2008
  Spectrally resolved energy dissipation rate and momentum flux of breaking
  waves. {\em J. Phys. Oceanogr.\/} {\bf 38}~(6), 1296--1312.

\bibitem[Grilli \& Svendsen(1990)]{Grilli1990}
{\sc Grilli, S.T. \& Svendsen, I.A.} 1990 Corner problems and global accuracy
  in the boundary element solution of nonlinear wave flows. {\em Engineering
  Analysis with Boundary Elements\/} {\bf 7}~(4), 178 -- 195.

\bibitem[Grilli {\em et~al.\/}(2001)Grilli, Guyenne \& Dias]{Grilli2001}
{\sc Grilli, St{\'e}phan~T., Guyenne, Philippe \& Dias, Fr{\'e}d{\'e}ric} 2001
  A fully non-linear model for three-dimensional overturning waves over an
  arbitrary bottom. {\em International Journal for Numerical Methods in
  Fluids\/} {\bf 35}~(7), 829--867.

\bibitem[Grilli \& Horrillo(1997)]{Grilli1997}
{\sc Grilli, St{\'e}phan~T. \& Horrillo, Juan} 1997 Numerical generation and
  absorption of fully nonlinear periodic waves. {\em Journal of Engineering
  Mechanics\/} {\bf 123}~(10), 1060--1069.

\bibitem[Grilli {\em et~al.\/}(1989)Grilli, Skourup \& Svendsen]{Grilli1989}
{\sc Grilli, S.~T., Skourup, J. \& Svendsen, I.~A.} 1989 An efficient boundary
  element method for nonlinear water waves. {\em Engineering Analysis with
  Boundary Elements\/} {\bf 6}~(2), 97 -- 107.

\bibitem[Grilli \& Subramanya(1994)]{Grilli1994}
{\sc Grilli, St{\'e}phan~T. \& Subramanya, Ravishankar} 1994 Quasi-singular
  integrals in the modeling of nonlinear water waves in shallow water. {\em
  Engineering Analysis with Boundary Elements\/} {\bf 13}~(2), 181 -- 191.

\bibitem[Grilli \& Subramanya(1996)]{Grilli1996}
{\sc Grilli, S.~T. \& Subramanya, R.} 1996 Numerical modeling of wave breaking
  induced by fixed or moving boundaries. {\em Computational Mechanics\/} {\bf
  17}~(6), 374 -- 391.

\bibitem[Grue {\em et~al.\/}(2003)Grue, Clamond, Huseby \& Jensen]{Grue2003}
{\sc Grue, John, Clamond, Didier, Huseby, Morten \& Jensen, Atle} 2003
  Kinematics of extreme waves in deep water. {\em Applied Ocean Research\/}
  {\bf 25}~(6), 355 -- 366.

\bibitem[Guyenne \& Grilli(2006)]{Guyenne2006}
{\sc Guyenne, P. \& Grilli, S.~T.} 2006 Numerical study of three-dimensional
  overturning waves in shallow water. {\em Journal of Fluid Mechanics\/} {\bf
  547}, 361--388.

\bibitem[Hayami(1990)]{Hayami1990}
{\sc Hayami, K} 1990 A robust numerical integration method for
  three-dimensional boundary element analysis. In {\em Bounday Elements XII:
  Applications in Stress Analysis, Potential and Diffusion\/} (ed. Honma
  Tanaka, Brebbia), {\em Bounday Elements XII\/}, vol.~1. Springer-Verlag.

\bibitem[Hayami(1991)]{Hayami1991}
{\sc Hayami, K} 1991 A projection transformation method for nearly singular
  surface boundary element integrals. PhD thesis, Computational mechanics
  institute, Wessex institute of technology, Southampton.

\bibitem[Hayami(2005{\natexlab{{\em a\/}}})]{Hayami2005}
{\sc Hayami, K.} 2005{\natexlab{{\em a\/}}} Variable transformations for nearly
  singular integrals in the boundary element method. {\em Tech. Rep.\/}. NII
  Technical Reports.

\bibitem[Hayami(2005{\natexlab{{\em b\/}}})]{Hayami2005a}
{\sc Hayami, K} 2005{\natexlab{{\em b\/}}} Variable transformations for nearly
  singular integrals in the boundary element method. {\em ublications of
  Research Institute for Mathematical Sciences\/} {\bf 41}, 821--842.

\bibitem[Hayami \& Matsumoto(1994)]{Hayami1994}
{\sc Hayami, Ken \& Matsumoto, Hideki} 1994 A numerical quadrature for nearly
  singular boundary element integrals. {\em Engineering Analysis with Boundary
  Elements\/} {\bf 13}~(2), 143 -- 154.

\bibitem[Hou \& Zhang(2002)]{Hou2002}
{\sc Hou, T.Y.a \& Zhang, P.b} 2002 Convergence of a boundary integral method
  for 3-d water waves. {\em Discrete and Continuous Dynamical Systems - Series
  B\/} {\bf 2}~(1), 1--34, cited By (since 1996)9.

\bibitem[Jessup \& Phadnis(2005)]{Jessup2005}
{\sc Jessup, A~T \& Phadnis, K~R} 2005 Measurement of the geometric and
  kinematic properties of microscale breaking waves from infrared imagery using
  a piv algorithm. {\em Measurement Science and Technology\/} {\bf 16}~(10),
  1961.

\bibitem[Johannessen(2010)]{Johannessen2010}
{\sc Johannessen, Thomas~B.} 2010 Calculations of kinematics underneath
  measured time histories of steep water waves. {\em Applied Ocean Research\/}
  {\bf 32}~(4), 391 -- 403.

\bibitem[Johannessen \& Swan(2001)]{Johannessen2001}
{\sc Johannessen, T.~B. \& Swan, C.} 2001 A laboratory study of the focusing of
  transient and directionally spread surface water waves. {\em Proceedings:
  Mathematical, Physical and Engineering Sciences\/} {\bf 457}~(2008), pp.
  971--1006.

\bibitem[Johannessen \& Swan(2003)]{Johannessen2003}
{\sc Johannessen, T.~B. \& Swan, C.} 2003 On the nonlinear dynamics of wave
  groups produced by the focusing of surface-water waves. {\em Proceedings of
  the Royal Society of London. Series A: Mathematical, Physical and Engineering
  Sciences\/} {\bf 459}~(2032), 1021--1052.

\bibitem[Katsardi \& Swan(2011)]{Katsardi2011}
{\sc Katsardi, V. \& Swan, C.} 2011 The evolution of large non-breaking waves
  in intermediate and shallow water. i. numerical calculations of
  uni-directional seas. {\em Proceedings of the Royal Society A: Mathematical,
  Physical and Engineering Science\/} {\bf 467}~(2127), 778--805.

\bibitem[Kinsman(1965)]{Kinsman1965}
{\sc Kinsman, Blair} 1965 {\em Wind waves : their generation and propagation on
  the ocean surface\/}. Englewood Cliffs, N.J.: Englewood Cliffs, N.J. :
  Prentice-Hall.

\bibitem[Kleiss \& Melville(2010)]{Kleiss2010a}
{\sc Kleiss, Jessica~M. \& Melville, W.~Kendall} 2010 Observations of wave
  breaking kinematics in fetch-limited seas. {\em J. Phys. Oceanogr.\/} {\bf
  40}~(12), 2575--2604.

\bibitem[Longuet-Higgins \& Stewart(1964)]{Longuet-Higgins1964}
{\sc Longuet-Higgins, M.S. \& Stewart, R.w.} 1964 Radiation stresses in water
  waves; a physical discussion, with applications. {\em Deep Sea Research and
  Oceanographic Abstracts\/} {\bf 11}~(4), 529 -- 562.

\bibitem[Longuet-Higgins(1987)]{Longuet-Higgins1987}
{\sc Longuet-Higgins, M.~S.} 1987 The propagation of short surface waves on
  longer gravity waves. {\em Journal of Fluid Mechanics\/} {\bf 177}, 293--306.

\bibitem[Longuet-Higgins \& Stewart(1960)]{Longuet-Higgins1960a}
{\sc Longuet-Higgins, M.~S. \& Stewart, R.~W.} 1960 Changes in the form of
  short gravity waves on long waves and tidal currents. {\em Journal of Fluid
  Mechanics\/} {\bf 8}, 565--583.

\bibitem[Ma(2010)]{Ma2010}
{\sc Ma, Qingwei.} 2010 {\em Advances in numerical simulation of nonlinear
  water waves\/}. Hackensack, NJ: World Scientific.

\bibitem[Ma {\em et~al.\/}(2001)Ma, Wu \& Eatock~Taylor]{Ma2001}
{\sc Ma, Q.~W., Wu, G.~X. \& Eatock~Taylor, R.} 2001 Finite element simulation
  of fully non-linear interaction between vertical cylinders and steep waves.
  part 1: methodology and numerical procedure. {\em International Journal for
  Numerical Methods in Fluids\/} {\bf 36}~(3), 265--285.

\bibitem[Melville(1983)]{Melville1983}
{\sc Melville, W.~K.} 1983 Wave modulation and breakdown. {\em Journal of Fluid
  Mechanics\/} {\bf 128}, 489--506.

\bibitem[Melville \& Matusov(2002)]{Melville2002}
{\sc Melville, W.~Kendall \& Matusov, Peter} 2002 Distribution of breaking
  waves at the ocean surface. {\em Nature\/} {\bf 417}~(6884), 58--63.

\bibitem[Melville {\em et~al.\/}(2002)Melville, Veron \& White]{MELVILLE2002a}
{\sc Melville, W.~Kendall, Veron, Fabrice \& White, Christopher~J.} 2002 The
  velocity field under breaking waves: coherent structures and turbulence. {\em
  Journal of Fluid Mechanics\/} {\bf 454}, 203--233.

\bibitem[Miller {\em et~al.\/}(1991)Miller, Shemdin \&
  Longuet-Higgins]{Miller1991}
{\sc Miller, Sarah~J., Shemdin, Omar~H. \& Longuet-Higgins, Michael~S.} 1991
  Laboratory measurements of modulation of short-wave slopes by long surface
  waves. {\em Journal of Fluid Mechanics\/} {\bf 233}, 389--404.

\bibitem[Nicholls(1998)]{Nicholls1998}
{\sc Nicholls, David~P.} 1998 Traveling water waves: Spectral continuation
  methods with parallel implementation. {\em Journal of Computational
  Physics\/} {\bf 143}~(1), 224 -- 240.

\bibitem[Park {\em et~al.\/}(2003)Park, Kim, Miyata \& Chun]{Park2003}
{\sc Park, J.~C., Kim, M.~H., Miyata, H. \& Chun, H.~H.} 2003 Fully nonlinear
  numerical wave tank (nwt) simulations and wave run-up prediction around 3-d
  structures. {\em Ocean Engineering\/} {\bf 30}~(15), 1969 -- 1996.

\bibitem[Phillips(1977)]{Phillips1977}
{\sc Phillips, Owen} 1977 {\em The dynamics of the upper ocean\/}, 2nd edn.
  Cambridge University Press Cambridge ; New York.

\bibitem[Phillips(1985)]{Phillips1985}
{\sc Phillips, O.~M.} 1985 Spectral and statistical properties of the
  equilibrium range in wind-generated gravity waves. {\em Journal of Fluid
  Mechanics\/} {\bf 156}, 505--531.

\bibitem[Rapp \& Melville(1990)]{Rapp1990}
{\sc Rapp, R.~J. \& Melville, W.~K.} 1990 Laboratory measurements of deep-water
  breaking waves. {\em Philosophical Transactions of the Royal Society of
  London. Series A, Mathematical and Physical Sciences\/} {\bf 331}~(1622),
  735--800.

\bibitem[Shemer(2013)]{Shemer2013}
{\sc Shemer, L.} 2013 On kinematics of very steep waves. {\em Natural Hazards
  and Earth System Science\/} {\bf 13}~(8), 2101--2107.

\bibitem[Song \& Banner(2002)]{Song2002}
{\sc Song, Jin-Bao \& Banner, Michael~L.} 2002 On determining the onset and
  strength of breaking for deep water waves. part i: Unforced irrotational wave
  groups. {\em Journal of Physical Oceanography\/} {\bf 32}~(9), 2541--2558.

\bibitem[Stansell \& MacFarlane(2002)]{Stansell2002}
{\sc Stansell, Paul \& MacFarlane, Colin} 2002 Experimental investigation of
  wave breaking criteria based on wave phase speeds. {\em Journal of Physical
  Oceanography\/} {\bf 32}~(5), 1269--1283.

\bibitem[Stokes(1847)]{Stokes1847}
{\sc Stokes, G.~G.} 1847 {On the theory of oscillatory waves}. {\em Trans.
  Camb. Phil. Soc.\/} {\bf 8}, 441--455.

\bibitem[Sutherland {\em et~al.\/}(1995)Sutherland, Greated \&
  Easson]{Sutherland1995}
{\sc Sutherland, J., Greated, C.A. \& Easson, W.J.} 1995 Variations in the
  crest kinematics of wave groups. {\em Applied Ocean Research\/} {\bf 17}~(1),
  55 -- 62.

\bibitem[Tayfun(1986)]{Tayfun1986}
{\sc Tayfun, M.~Aziz} 1986 On narrow-band representation of ocean waves: 1.
  theory. {\em Journal of Geophysical Research: Oceans\/} {\bf 91}~(C6),
  7743--7752.

\bibitem[Telles(1987)]{Telles1987}
{\sc Telles, J.C.F} 1987 A self-adaptive co-ordinate transformation for
  efficient numerical evaluation of general boundary element integrals. {\em
  Int. J. Numer. Meth. Engng.\/} {\bf 24}~(5), 959--973.

\bibitem[Viotti {\em et~al.\/}(2014)Viotti, Carbone \& Dias]{Viotti2014}
{\sc Viotti, Claudio, Carbone, Francesco \& Dias, Frédéric} 2014 Conditions for
  extreme wave runup on a vertical barrier by nonlinear dispersion. {\em
  Journal of Fluid Mechanics\/} {\bf 748}, 768--788.

\bibitem[West {\em et~al.\/}(1987)West, Brueckner, Janda, Milder \&
  Milton]{West1987}
{\sc West, Bruce~J., Brueckner, Keith~A., Janda, Ralph~S., Milder, D.~Michael
  \& Milton, Robert~L.} 1987 A new numerical method for surface hydrodynamics.
  {\em Journal of Geophysical Research: Oceans\/} {\bf 92}~(C11), 11803--11824.

\bibitem[Wiegel(1964)]{Wiegel1964}
{\sc Wiegel, R.~L.} 1964 {\em Oceanographical Engineering\/}. Englewood Cliffs,
  N.J.: Englewood Cliffs, N.J., Prentice-Hall.

\bibitem[Xu \& Guyenne(2009)]{Xu2009}
{\sc Xu, Liwei \& Guyenne, Philippe} 2009 Numerical simulation of
  three-dimensional nonlinear water waves. {\em Journal of Computational
  Physics\/} {\bf 228}~(22), 8446 -- 8466.

\bibitem[Xue {\em et~al.\/}(2001)Xue, X\&uuml;, Liu \& Yue]{XUE2001}
{\sc Xue, Ming, X\&uuml;, Hongbo, Liu, Yuming \& Yue, Dick K.~P.} 2001
  Computations of fully nonlinear three-dimensional wavebody interactions. part
  1. dynamics of steep three-dimensional waves. {\em Journal of Fluid
  Mechanics\/} {\bf 438}~(-1), 11--39.

\bibitem[Zakharov(1968)]{Zakharov1968}
{\sc Zakharov, V.E.} 1968 Stability of periodic waves of finite amplitude on
  the surface of a deep fluid. {\em Sov. Phys. J. Appl. Mech. Tech. Phys.\/}
  {\bf 4}, 190--194.

\end{thebibliography}

\end{document}